\input{aipcheck}

\documentclass[
    ,final            
  ]
  {aipproc}

\layoutstyle{6x9}

\begin{document}

\title{Normal Gas-Rich Galaxies in the Far-Infrared: \\
The Legacy of ISOPHOT}

\author{Richard J. Tuffs \& Cristina C. Popescu}{
  address={Max Planck Institut f\"ur Kernphysik, Astrophysics Department,\\ 
Saupfercheckweg 1, D-69117 Heidelberg,\\ e-mail:Richard.Tuffs@mpi-hd.mpg.de; Cristina.Popescu@mpi-hd.mpg.de}
}

\begin{abstract}
Following on from IRAS, the ISOPHOT instrument on board the Infrared Space
Observatory (ISO) has provided a huge advancement
in our knowledge of the phenomenology of the far-infrared (FIR) emission of
normal galaxies and the underlying physical processes. Highlights include:
the discovery of an extended cold dust emission component, present in all 
types of gas-rich galaxies and carrying the bulk of the dust luminosity; 
the definitive characterisation of the spectral energy distribution
in the FIR, revealing the channels through which stars power
the FIR light; the derivation of realistic 
geometries for stars and dust from ISOPHOT imaging and the discovery of cold 
dust associated with HI extending beyond the optical body of galaxies.
\end{abstract}
\maketitle

\section{1. Introduction}

Whereas IRAS provided the first systematic survey of far-infrared (FIR) 
emission from normal galaxies, it has been the photometric, imaging and 
spectroscopic capabilities of the ISOPHOT instrument (Lemke et
al. \cite{lem96}) on board the Infrared 
Space Observatory (ISO; Kessler et al. \cite{kes96}) 
which have unravelled the basic physical processes 
giving rise to this FIR emission. Thanks to the broad spectral
grasp of ISOPHOT, the bulk of the emission from dust could be measured, 
providing the first quantitative assessment of the fraction of stellar 
light re-radiated by dust. The battery of filters has led to a
definitive characterisation of the spectral energy distribution (SED)
in the FIR, revealing the contribution of the different stellar
populations in powering the FIR emission. 
The imaging capabilities have unveiled the complex morphology of
galaxies in the FIR, and their changing appearance with FIR 
wavelength. They also allowed the exploration of hitherto undetected
faint diffuse regions of galaxies. 

In this review we will concentrate on the
FIR properties of normal nearby galaxies. By normal
we essentially mean that their SEDs are not powered by accretion.  
We will begin with spiral galaxies and then we 
will move to the other class of gas-rich galaxies, the dwarfs. 

\section{2. Spiral Galaxies}
\subsection{2.1 Spatial distributions}
\label{ssec:spatdist}
\subsubsection{2.1.1 FIR Morphologies}  
\label{sssec:firmorph}

ISOPHOT imaged three nearby galaxies (M~31: Haas et al. \cite{haa98a}; M~33:
Hippelein et al. \cite{hip03} and M~101: Tuffs \& Gabriel \cite{tg03}) in the 60 to
200\,${\mu}$m range, with sufficient linear resolution to easily
distinguish between the main morphological components in the FIR -
nucleus, spiral arms and underlying disk.  The main discovery, made
possible by the unprecedented surface brightness sensitivity longwards
of 100\,${\mu}$m, was the existence of large amounts of cold dust
associated both with the spiral arms and with the underlying
disk. This dust was too cold to have been seen by IRAS. Furthermore,
ground-based submillimeter (submm) facilities have lacked the surface 
brightness sensitivity to map the diffuse component of the cold dust 
associated with the underlying disk. Recently, however, a submillimeter
counterpart of the diffuse disk known from FIR studies was revealed through
deep SCUBA mapping of M~51 (Meijerink et al. \cite{mei04}).

\begin{figure} 
\includegraphics[height=.3\textheight]{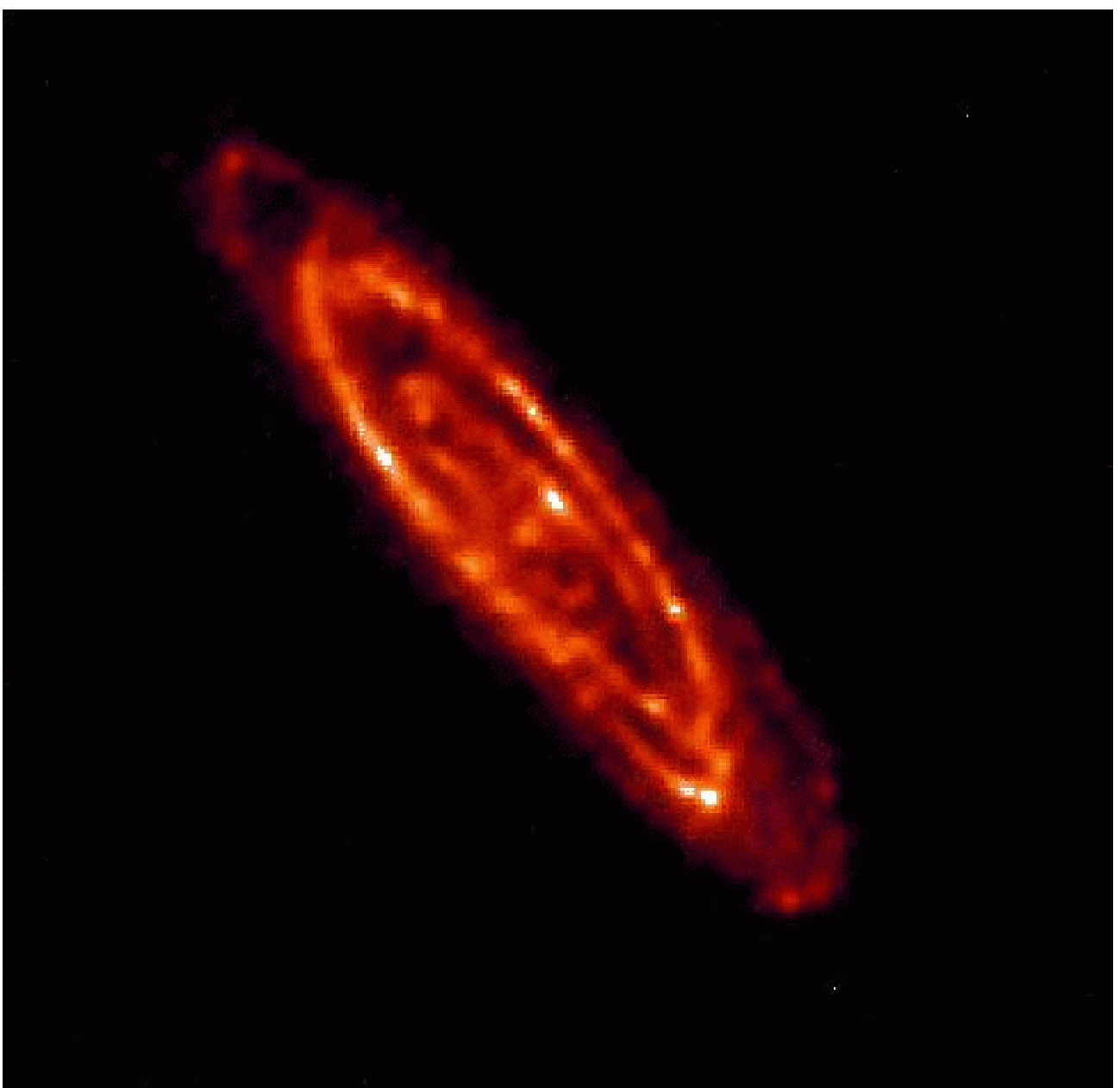}
\includegraphics[height=.25\textheight]{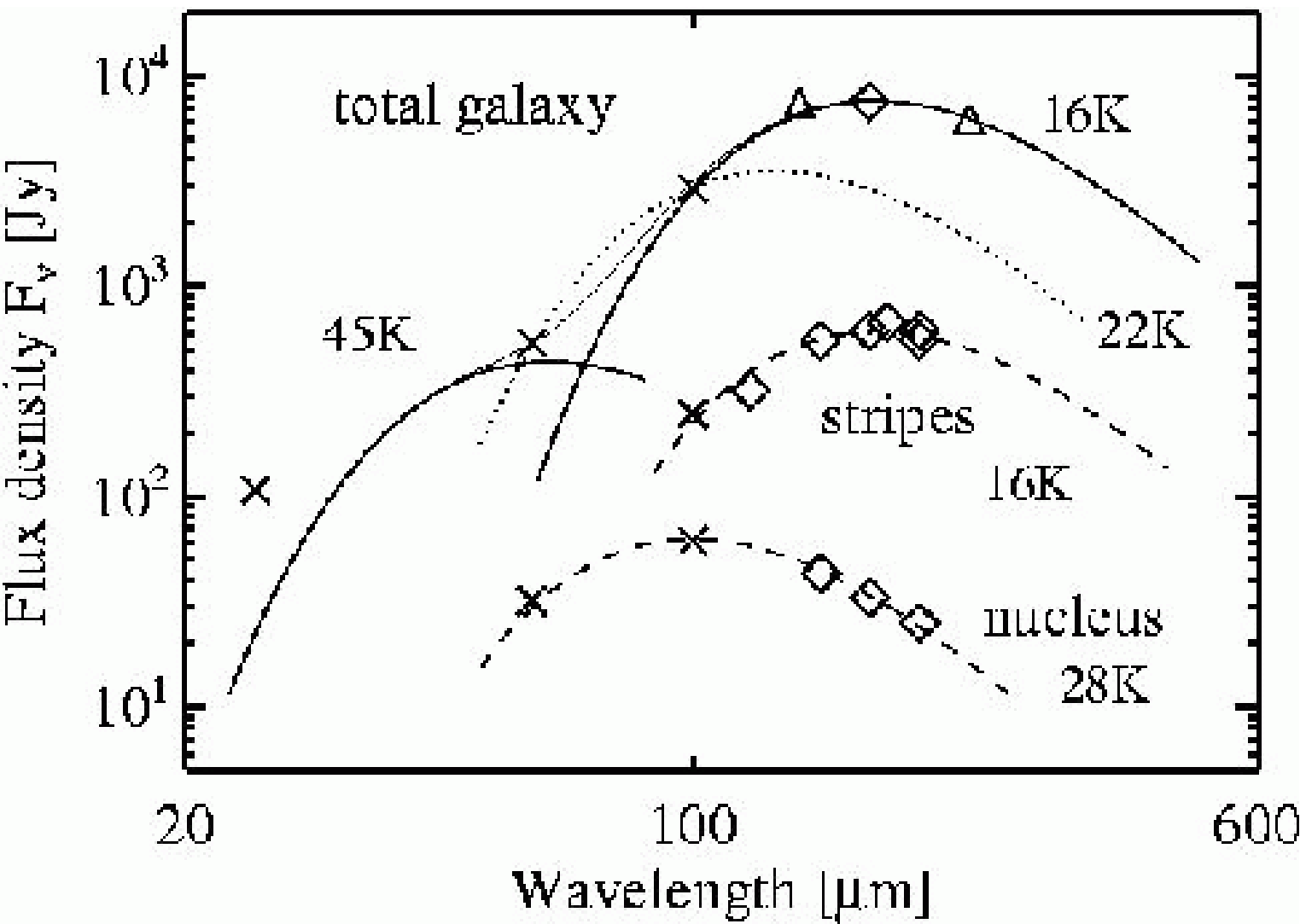}
\caption{Left: ISOPHOT 170$\,{\mu}$m map of M~31 (Haas et al. \cite{haa98a}),
with an angular resolution of 1.3$^{\prime}$. North is towards the top,
and East is towards the left. The field size is $2.9 \times 2.9$
degrees. Right: Infrared SED of M~31 (Haas et
al. \cite{haa98a}). The data are shown by symbols (diamonds ISO, crosses IRAS,
triangles DIRBE) with the size being larger than the errors. The
blackbody curves with emissivity proportional to ${\lambda}^{-2}$ are
shown by lines. The dotted line with T=22\,K through the IRAS 60 and
100\,${\mu}$m data points indicates what one would extrapolate from
this wavelength range alone without any further assumptions.}
\end{figure}

In the case of the Sab galaxy M~31, most of the emission at
170\,${\mu}$m arises from the underlying disk, which has a completely
diffuse appearance (Fig.~1, left panel). This diffuse disk
emission can be traced out to a radius of 22\,kpc, so the galaxy has a
similar overall size in the FIR as seen in the optical bands.
Fig.~1 (left panel) also shows that at 170\,${\mu}$m the
spiral arm component is dominated by a ring of 10\,kpc radius.  In
addition, there is a faint nuclear source, which is seen more
prominently in HIRES IRAS 60\,${\mu}$m maps at similar resolution and
in H${\alpha}$.  The overall SED (Fig.~1, right panel)
can be
well described as a superposition of two modified ($\beta$=2) Planck
curves, with dust temperatures $T_{\rm D}$ of 16 and 45\,K.  The cold
dust component at 16\,K arises from both the ring structure (30$\%$)
and the diffuse disk (70$\%$; Haas, private communication),
illustrating the importance of the diffuse emission at least for this
example.  The 45\,K component matches up well with HII regions within
the star-formation complexes in the ring structure. Associated with
each star-formation complex are also compact, cold emission sources
(Schmitobreick et al. \cite{smi00}) with dust
temperatures in the 15 to 20\,K range.  These could well represent the
parent molecular clouds in the star-formation complexes which gave
rise to the HII regions.  Detailed examination of the morphology of
the ring shows a smooth component of cold dust emission as well as the
discrete cold dust sources. Finally, the nuclear emission was fitted
by a 28\,K dust component.

The ISOPHOT maps of the Sc galaxies M~33 and M~101 show the same
morphological components as seen in M~31, with the difference that the
spiral arm structure can be better defined in these later-type
spirals. Also the star-formation complexes in the spiral arms show
similar SEDs to those seen in M~31.

In conclusion, the characteristics of the FIR emission from the main
morphological components of spiral galaxies are:

\begin{itemize}

\item {\bf nucleus}: an unresolved warm source with $T_{\rm
D}\,\sim\,$30\,K

\item {\bf spiral arms}: a superposition of:

\begin{itemize}

\item localised warm emission with 40\,$\le\,T_{\rm D}\,\le\,$60\,K
from HII regions

\item localised cold emission with 15\,$\le\,T_{\rm D}\,\le\,$20\,K
from parent molecular clouds

\item diffuse emission running along the arms

\end{itemize}

\item {\bf disk}: an underlying diffuse (predominantly cold) emission
with 12\,$\le\,T_{\rm D}\,\le\,$20\,K

\end{itemize}

\subsubsection{2.1.2 The extent of spiral disks in the FIR}
\label{sssec:firextent}

 Further information about the distribution of dust in spiral
disks is provided by FIR observations of galaxies more distant than the
highly resolved local galaxies discussed in Sect.~2.1.1,
but still close enough to resolve the diffuse disk at the longest FIR
wavelengths accessible to ISO. In a study of eight spiral galaxies
mapped by ISOPHOT at 200\,${\mu}$m, Alton et al. (\cite{alt98b}; see also
Davies et al. \cite{dav99} for NGC~6946) showed that the observed scalelength
of FIR emission at 200\,${\mu}$m is greater than that found by IRAS at
60 and 100\,${\mu}$m.  Thus, the scalelength of the FIR emission
increases with increasing FIR wavelength. This result was reinforced
using LWS measurements of the dust continuum by Trewhella et
al. \cite{tre00}, and can also be inferred from Hippelein et
al. \cite{hip03} for M~33. This implies that the bulk of
the 200\,${\mu}$m emission arises from grains heated by a radially
decreasing radiation field, as would be expected for grains in the
diffuse disk. If most of the 200\,${\mu}$m emission had arisen from
localised sources associated with the parent molecular clouds, there should 
be no FIR colour gradient in the galaxy,
since the SEDs of sources which are locally heated should not depend 
strongly on position.

\begin{figure}[htb] 
\includegraphics[height=.28\textheight]{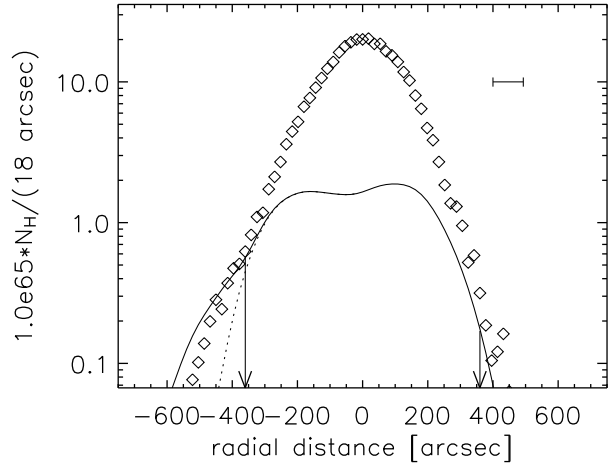} 
\caption{The radial profiles of HI emission (from Swaters et al. \cite{swa97}) 
 convolved with the
  ISOPHOT PSF (solid line) and of $200\,{\mu}$m FIR emission (symbols)
  of NGC~891 (Popescu \& Tuffs \cite{poptuf03}). Note that the extent and asymmetry of the
  200\,${\mu}$m\ emission follow that of the HI emission. 
  The profiles are sampled at
  intervals of
  18$^{\prime\prime}$. The negative radii correspond to the southern side of 
  the
  galaxy and the galaxy was scanned at 60 degrees with respect to the major
  axis. The units of the FIR profile are W/Hz/pixel, multiplied by a
  factor of $2\times 10^{-22}$ and the error bars are smaller than the 
  symbols. The horizontal bar delineates the FWHM of the ISOPHOT 
  PSF of 93$^{\prime\prime}$. 
  The vertical arrows indicate the maximum extent of the optically
  emitting disk. The dotted line represents a modified HI profile 
  obtained in the southern side from the original one by cutting
  off its emission at the edge of the optical disk and by convolving
  it with the ISOPHOT PSF.}

\label{fig:f2}
\end{figure}

The second result to come out of the studies by Alton et al. \cite{alt98a}
and Davies et al. \cite{dav99} is that the observed scalelength at
200\,${\mu}$m is comparable to or exceeds the scalelength of the
optical emission (see also Tuffs et al. \cite{tuf96}). As noted by 
Alton et al. \cite{alt98a}, this result implies that
the {\it intrinsic} scalelength of the dust in galaxies is greater
than that of the stars. This is because the apparent scalelength of
stars should increase with increasing disk opacity (since the inner
disk is expected to be more opaque than the outer disk) whereas the
apparent scalelength of the dust emission will be less than the
intrinsic scalelength (due to the decrease in grain temperature with
increasing galactocentric radius). 
The extraction of the precise relation between the
intrinsic scalelengths of stars and dust requires a self-consistent calculation
of the transfer of radiation through the disk (see Popescu \&
Tuffs \cite{poptuf05}).
The reason for the difference between the intrinsic scalelength of stars and
dust in galaxies is not self-evident, since
it is the stars themselves which are thought to be the sources of
interstellar grains (produced either in the winds of evolved
intermediate mass stars or perhaps in supernovae). One might speculate
either that there is a mechanism to transport grains from the inner
disk to the outer disk, or that the typical lifetimes of grains
against destruction by shocks is longer in the outer disk than it is
in the inner disk.

While Alton et al. and Davies et al. showed that the scalelength of
the 200\,${\mu}$m emission was comparable to or slightly larger than
that of the optical emission, these studies did not actually detect
grain emission beyond the edge of the optical disk. Since spiral
galaxies in the local universe are commonly observed to be embedded in
extended disks of neutral hydrogen - the so called ``extended HI
disks'', it is a natural question to ask whether these gaseous disks
contain grains. 
This question was answered in the affirmative by
Popescu \& Tuffs \cite{poptuf03}, through dedicated deep FIR maps of a large
field encompassing the entire HI disk of the edge-on spiral galaxy
NGC~891, made using ISOPHOT at 170 and 200\,${\mu}$m (see Fig.~2.). 

\begin{figure}[htb] 
\includegraphics[height=.3\textheight]{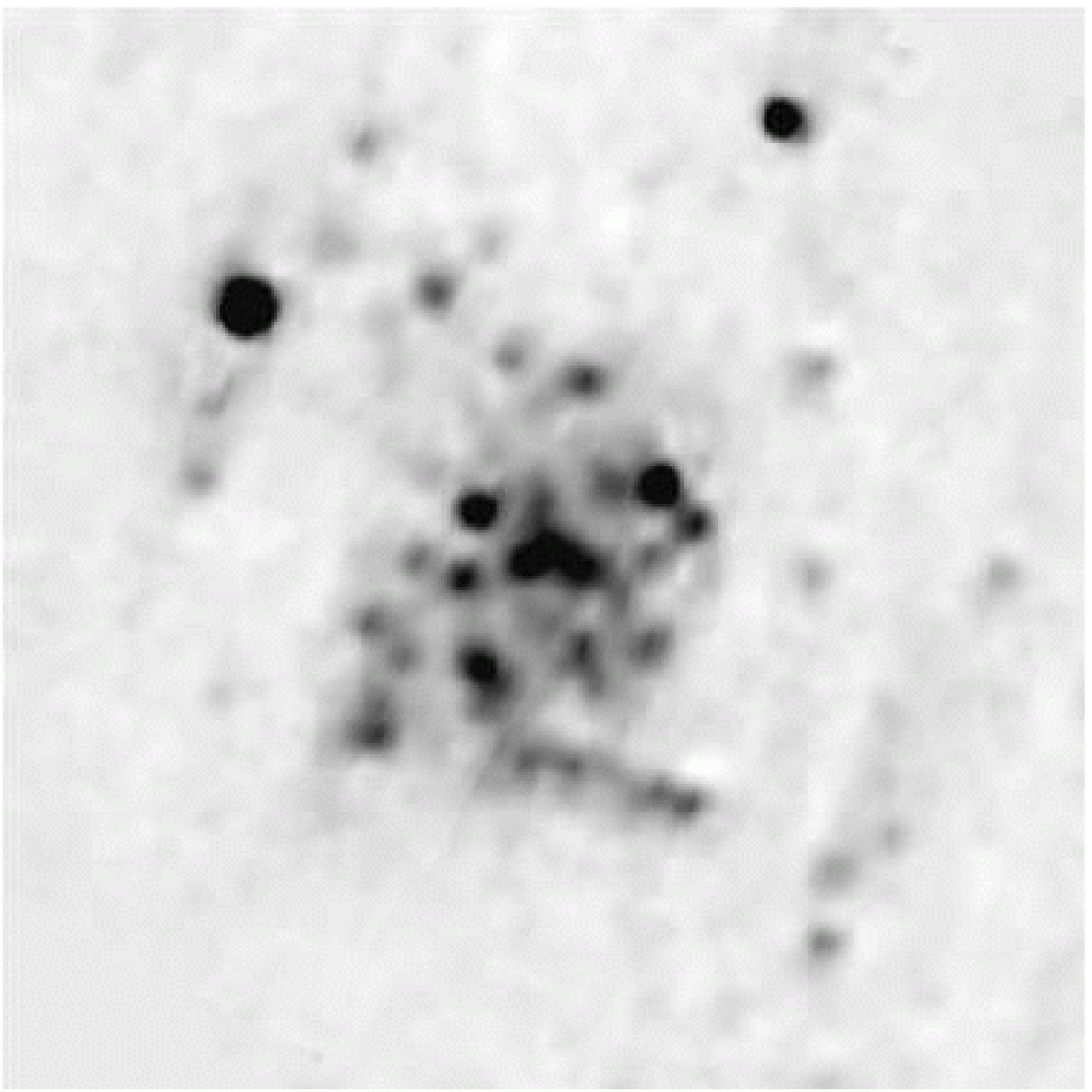}
\includegraphics[height=.3\textheight]{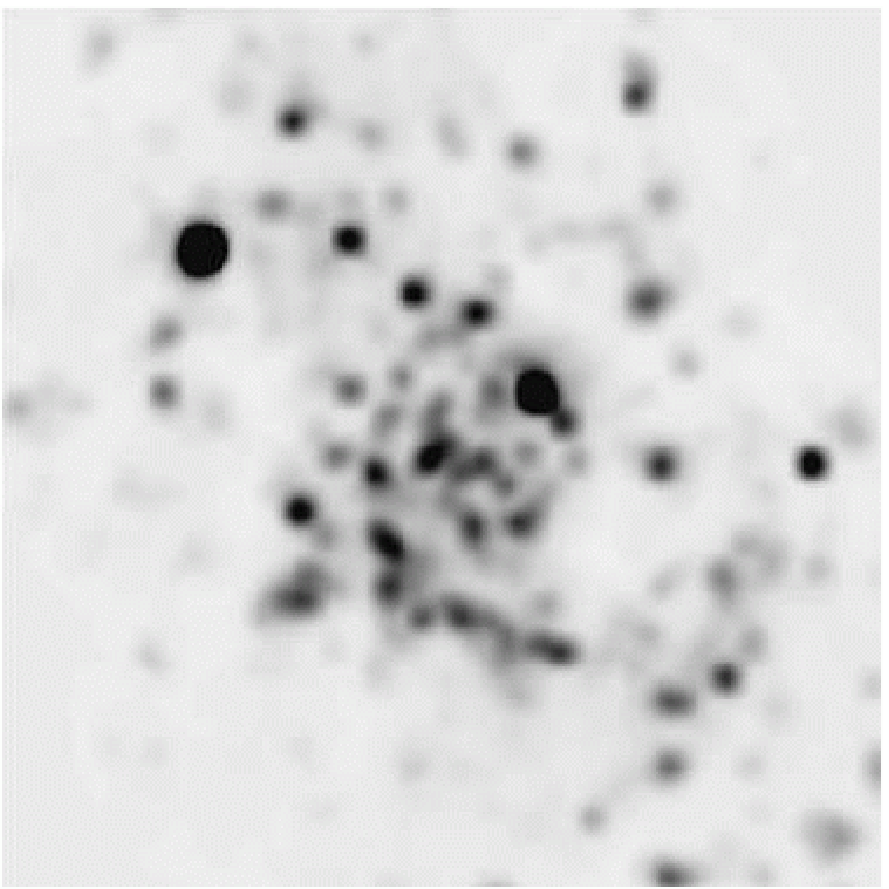}

\caption{Left: Distribution of the localised warm dust component at
  60\,${\mu}$m, $F^{\rm l}_{60}$, in M~33
  (Hippelein et al. \cite{hip03}). This is the scaled difference map
  $2(F_{60}-0.165\times F_{160}$), with the factor 0.165 given by the average
  flux density ratio $F_{60}/F_{170}$ in the interarm regions. Right:
  $H{\alpha}$ map of M~33 convolved to a resolution of 60$^{\prime\prime}$.}

  \label{fig:f3}
\end{figure}

The large amounts of grains found in the extended HI disk of NGC~891
(gas-to-dust ratio of $\sim 1\%$) clearly shows that this gaseous disk
is not primordial, left over from the epoch of galaxy formation. It
was suggested that the detected grains could have either been
transported from the optical disk (via the halo, using mechanisms such
as those proposed by Ferrara \cite{fer91}, Davies et al. \cite{dav98}, 
Popescu et al. \cite{pop00a} or through the action of macro turbulence) or that they could
have been produced outside the galaxy (for example transferred in
interactions with other galaxies). It is interesting to note that,
although the dust emission is seen towards the HI component, the
grains may not actually be embedded in the neutral ISM. Instead, this
dust could trace an ``unseen'' molecular component, as proposed by
Pfenniger \& Combes \cite{pc94}, Pfenniger, Combes \& Martinet \cite{pfe94},
Gerhard \& Silk \cite{gs96}, and Valentijn et al. \cite{val99b}.  This cold 
molecular gas component has been invoked as a dark matter component to 
explain the
flat rotation curves of spiral galaxies. Its presence might also
reconcile the apparent discrepancy between the very low metallicities
measured in HII regions in the outer disk (Ferguson, Gallagher \&
Wyse \cite{ferg98}) and the high ratio of dust to gas (on the assumption that all
gas is in form of HI) found by Popescu \&
Tuffs \cite{poptuf03} in the extended HI disk of NGC~891.

\subsubsection{2.1.3 Comparison with morphologies at other wavelengths}
\label{sssec:compwave}

\noindent
{\it Comparison with H${\alpha}$}

\begin{figure}[htb] 
\includegraphics[height=.3\textheight]{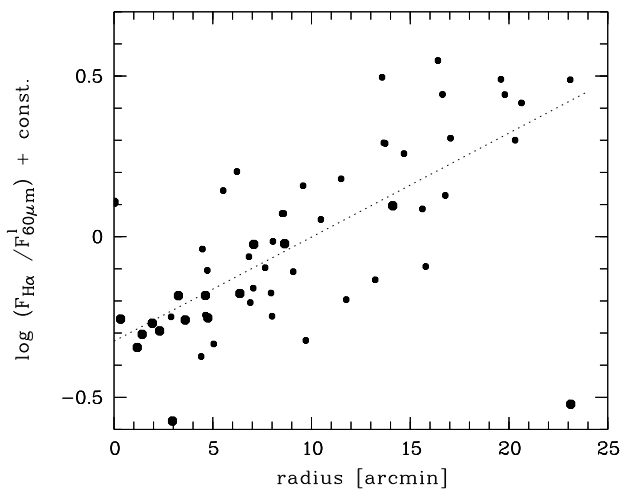}

\caption{The ratio of $F_{H\alpha}$ to the localised warm dust
component at 60\,${\mu}$m, $F^{\rm l}_{60}$, for the star-forming
regions in M~33 (Hippelein et al. \cite{hip03}) versus distance from the
galaxy centre. Symbol sizes indicate the brightness of the sources.}

\label{fig:f4}
\end{figure} 

\noindent
For the highly resolved galaxy M~33, Hippelein et al. \cite{hip03} showed
that there is a strong resemblance between the morphology of the
localised warm dust component at 60\,${\mu}$m ($F^{\rm l}_{60}$; obtained at
each direction by subtracting the 170\,${\mu}$m\ map scaled by the ratio of the
60/170\,${\mu}$m\ brightness in the interarm regions) and
the morphology of the H$\alpha$ emission (see Fig.~3),
indicating that the 60\,${\mu}$m localised emission traces the
star-formation complexes.  The $F_{H\alpha}/F^{\rm l}_{60}$ ratio (see
Fig.~4) for the star-formation complexes shows a clear
systematic increase with increasing radial distance from the centre
(allowing for the [NII] line contribution decreasing with distance,
the slope would be even steeper).  Very probably this is due to the presence 
of a larger scale gradient of opacity affecting the recombination line
fluxes. 

\vspace{0.5cm}
\noindent
{\it Comparison with UV}

A fundamental property of spiral galaxies is the fraction of light
from young stars which is re-radiated by dust. This property can be
investigated as a function of position in the galaxy by a direct
comparison of ISOPHOT maps at 60, 100 and 170\,${\mu}$m with UV maps
obtained with GALEX (Galaxy Evolution Explorer; Martin et al. \cite{mar05}) in
its near-UV (NUV; 2310\,\AA) and far-UV (FUV; 1530\,\AA) bands. Such a
comparison was performed for M~101 by Popescu et al. \cite{pop05}.  The top
panels in Fig.~5 display the 100\,${\mu}$m ISOPHOT image
(left) together with the corresponding ``total UV'' (integrated from
1412 to 2718\,\AA) image (right).  
Comparison between the ratio image
(100\,${\mu}$m/UV) (bottom left panel) and an image of the ``spiral
arm fraction'' (the fraction of the UV emission from the spiral arm within an
ISOPHOT beam; bottom right panel) shows that the high values of the
100\,${\mu}$m/UV ratio trace the interarm regions. In other words the
``spiral features'' in the ratio image are in reality regions of
diffuse emission which are interspaced with the real spiral features,
as seen in the ``spiral arm fraction'' image.  

\begin{figure}[htb] 
\includegraphics[scale=0.8]{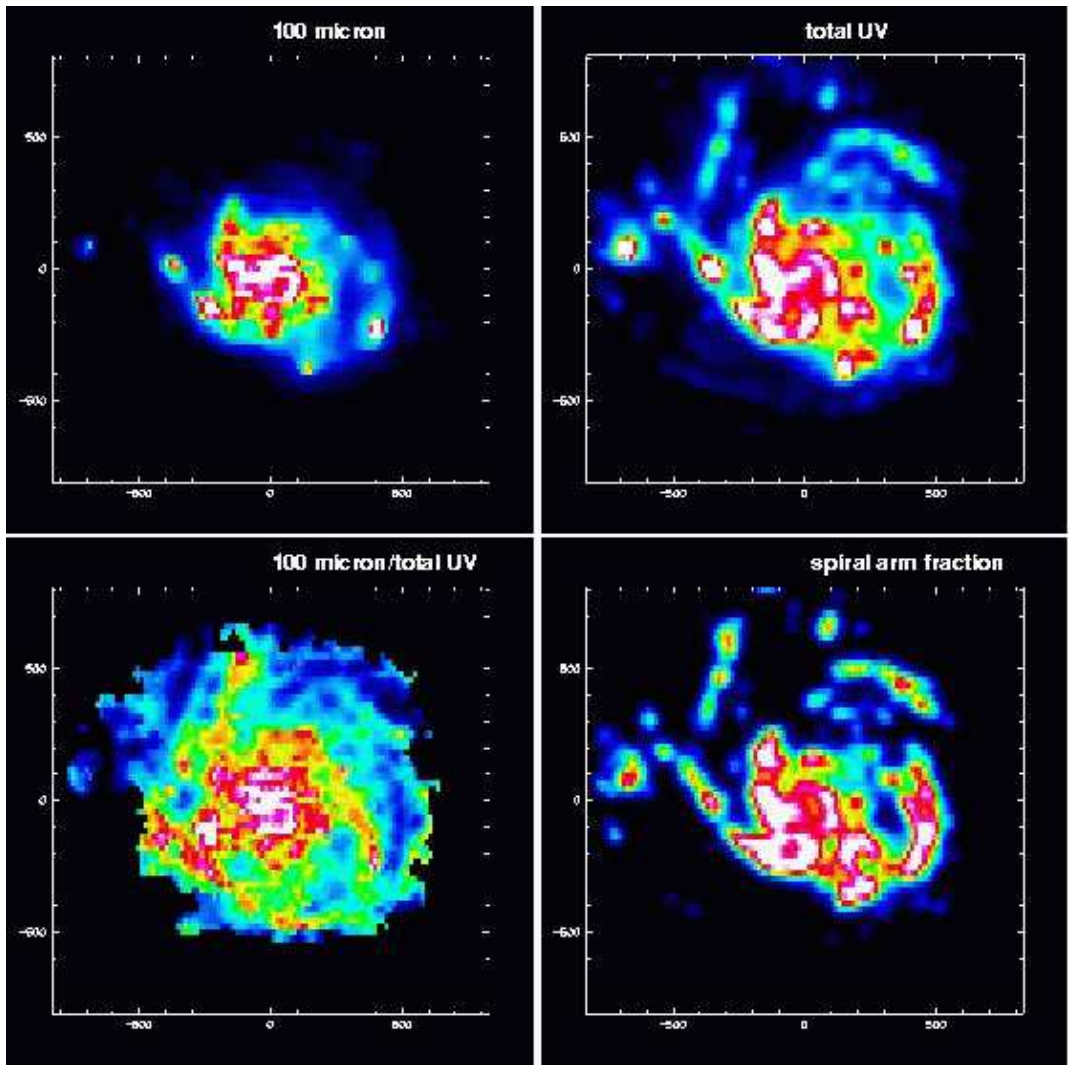} 
\caption{
FIR-UV comparison for M~101 (Popescu et al. \cite{pop05}). Top left:
filter-integrated 100\,${\mu}$m ISOPHOT image.  Top right: ``total UV''
image. Bottom left: ratio of the filter-integrated 100\,${\mu}$m ISOPHOT image
divided by the corresponding ``total UV'' image. Bottom right:
the ``spiral arm fraction''.  All
panels depict a field of 27.7$^{\prime}\times 27.1^{\prime}$ centered
at $\alpha^{2000}=14^{\rm h}03^{\rm m}13.11^{\rm s}$;
$\delta^{2000}=54^{\circ}21^{\prime}06.6^{\prime\prime}$, and have the
orientation, resolution and sampling ($15.33^{\prime\prime}\times
23.00^{\prime\prime}$) of the 100\,${\mu}$m ISOPHOT image.} 

\label{fig:f5} 
\end{figure} 

\begin{figure}[htb] 
\includegraphics[scale=0.6]{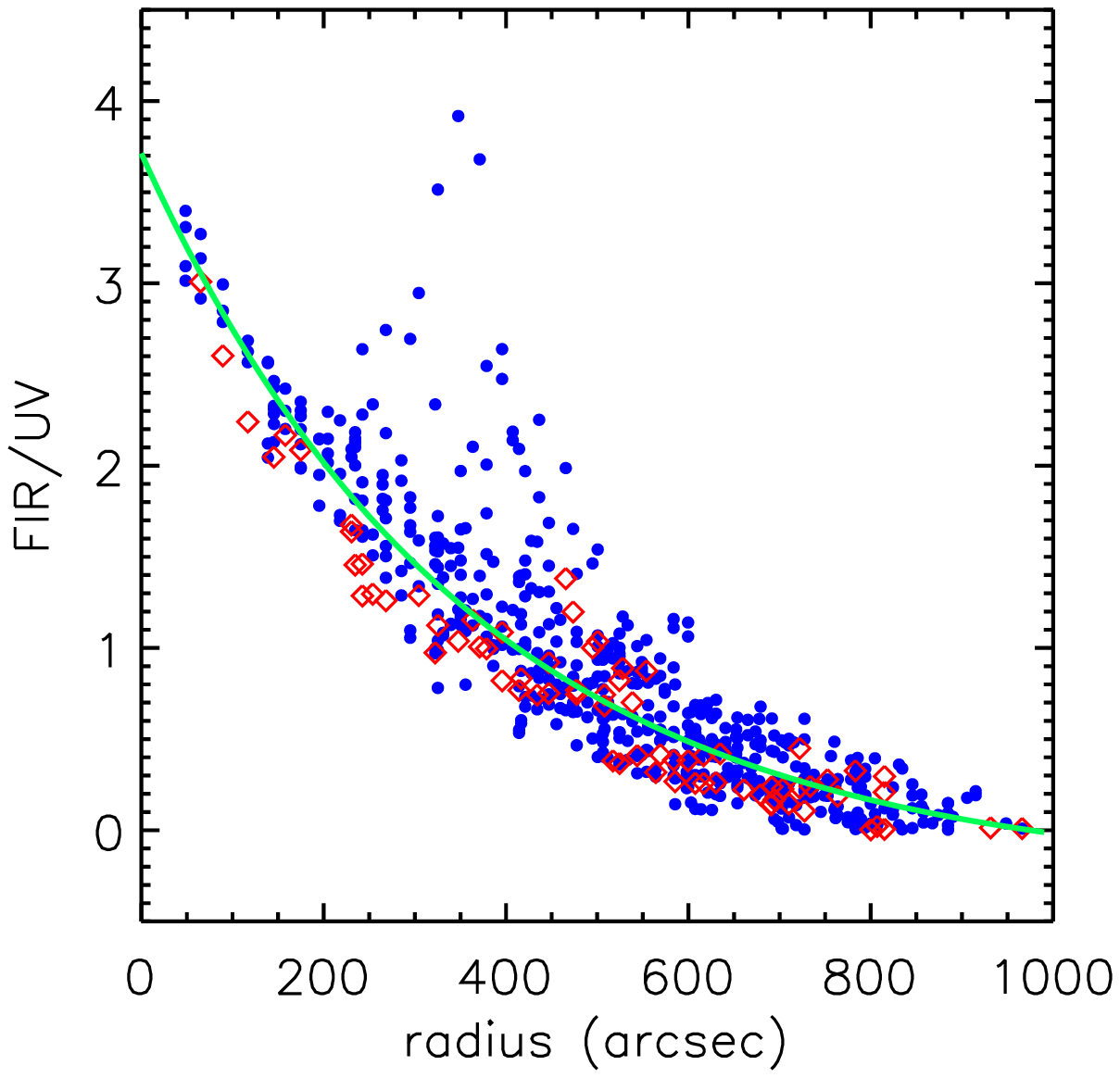} 

\caption{
The pixel values of the FIR/UV ratio map of M~101 (Popescu et
al. \cite{pop05}) at the resolution of the 170\,${\mu}$m image versus angular
radius. The blue dots are for lines of sight towards interarm regions
and the red diamonds towards the spiral arm regions. The green solid
line is an offset exponential fit to the data.}

\label{fig:f6}
\end{figure}

The trend for the FIR/UV ratio to be higher in the
diffuse interarm regions than in the spiral-arms is seen in
Fig.~6 from the segregation of the blue dots and red
diamonds at a given radius. This apparently surprising result was
explained in terms of the escape probability of UV photons from spiral
arms and their subsequent scattering in the interarm regions, and in
terms of the larger relative contribution of optical photons to the
heating of the dust in the interarm regions. The combined effect of
the optical heating and the scattering of the UV emission means that
the FIR/UV ratio will not be a good indicator of extinction in the
interarm region.

Despite these local variations, the main result of Popescu et
al. \cite{pop05} is the discovery of a tight dependence of the FIR/UV ratio
on radius, with values monotonically decreasing from $\sim 4$ in the
nuclear region to nearly zero towards the edge of the optical disk
(see Fig.~6).  This was interpreted in terms of the
presence of a large-scale distribution of diffuse dust having a
face-on optical depth which decreases with radius and which dominates
over the more localised variations in opacity between the arm and
interarm regions.

\subsection{2.2 Integrated properties}
\label{ssec:int}

The overriding result of all ISOPHOT studies of the integrated properties
of normal galaxies in the FIR is that their SEDs in the
40-200\,${\mu}$m spectral range require both warm and cold dust
emission components to be fitted. Although the concept of warm and cold 
emission components is as old as IRAS (de Jong et al. \cite{dej84}), it only became 
possible to directly measure and 
spectrally separate these components using ISOPHOT's multi-filter coverage
of the FIR regime out to 200\,${\mu}$m. 

In order to investigate the integrated properties of local universe gas-rich
galaxies, a number of statistical samples were constructed.
All these projects were complementary in terms of selection and observational
goals. In descending order of depth (measured in terms of a typical 
bolometric luminosity of the detected objects), the published surveys are:

{\bf The ISOPHOT Virgo Cluster Deep Survey} (IVCDS; Tuffs et
al. \cite{tuf02a},\cite{tuf02b}, Popescu et al. \cite{pop02}) represents 
the {\it deepest survey} (both in
luminosity and surface brightness terms) of normal galaxies
measured in the FIR with ISOPHOT. A complete volume- and luminosity-limited 
sample of 63
gas-rich Virgo Cluster galaxies selected from the Virgo Cluster
Catalogue (Binggeli et al. \cite{bin85}; see also Binggeli et al. 
\cite{bin93}) with
Hubble types later than S0 and brighter than $B_{\rm T} \le 16.8$ were
mapped with ISOPHOT at 60, 100 and 170\,${\mu}$m. 
The IVCDS sample was (in part) also observed with the LWS (Leech et
al. \cite{lee99}) and with ISOCAM (Boselli et al. 
\cite{bos03b}, see also Boselli et al. \cite{bos97}).

The IVCDS provides a database for statistical 
investigations of the FIR SEDs of gas-rich 
galaxies in the local universe spanning a broad range in star-formation 
activity and morphological types, including dwarf systems and galaxies with 
rather quiescent star-formation activity.

{\bf The Coma/A1367 Survey} Contursi et al. \cite{con01} consists of 6 spiral 
and 12
irregular galaxies having IRAS detections at 60$\,{\mu}$m. The
galaxies were selected to be located within 2 or 1 degrees of the
X-ray centres of Coma and A1367 clusters, respectively, with emphasis
on peculiar optical morphologies. Each galaxy was observed in a single
pointing with ISOPHOT, 
at 120, 170 and 200$\,{\mu}$m, as well
as mapped with ISOCAM in the 6.75 and 15\,${\mu}$m\ broadband filters. The sample 
provides a database of integrated flux densities for a
pure cluster sample of high luminosity spiral and irregular galaxies.

{\bf The ISO Bright Spiral Galaxies Survey} (Bendo et al. \cite{ben02a},
\cite{ben02b}, \cite{ben03}) consists of 77 spiral and S0
galaxies chosen from the Revised Shapley-Ames Catalog (RSA), with
$B_{\rm T}\,\le\,12.0$. Almost all are IRAS sources. Mainly an ISOCAM
mapping survey with the 12\,${\mu}$m\ filter, the
project also used ISOPHOT to take 60, 100 and 170$\,{\mu}$m short
stares towards the nucleus of the galaxies and towards background
fields. The sample provides a database of MIR morphologies and FIR surface 
brightnesses of the central regions of bright spiral galaxies, including S0s.

The ISO Bright Spiral Galaxies Survey and the IVCDS 
represent
the principle investigations of optically selected samples of normal
galaxies. It should be emphasised that the main difference between
them is primarily one of shallow versus deep, rather than field versus
cluster, since by design the Virgo Sample predominantly consists of
infalling galaxies from the field, and no cluster specific effects
could be found (see also Contursi et al. \cite{con01} for Coma/A1367 Sample).

{\bf The ISO Key Project on Normal Galaxies Sample} 
consists of 69 galaxies selected to span the whole range of the
classical IRAS colour-colour diagram (Helou \cite{hel86}). Since IRAS detected a
vast number of galaxies in its four bands, the selection was also made
to span the Hubble sequence evenly and provide a broad range of
IR luminosities, dust temperatures (as determined by IRAS) and
in star-formation activity. Single pointing toward the centres of the galaxies 
in the sample were made by ISOPHOT-S (Lu et al. \cite{lu03}) to measure the
2.5-12\,${\mu}$m\ spectra and by ISOPHOT-C at 170\,${\mu}$m. The galaxies from
this sample were also mapped with ISOCAM (Dale et al. \cite{dal00}) and their 
main cooling lines in the 
FIR were measured with ISOLWS (Malhotra et al. \cite{mal01}).


{\bf The ISOPHOT Serendipity Survey} (Stickel et al. \cite{sti00}) has initially
catalogued 115 galaxies with $S_{\nu}\,\ge\,2$\,Jy at 170$\,{\mu}$m
and with morphological types predominantly S0/a\,-\,Scd. This sample
provides a database of integrated 170$\,{\mu}$m flux densities for
relatively high luminosity spiral galaxies, all detected by IRAS at 60
\& 100$\,{\mu}$m. Recently a catalogue of 1900 galaxies was released
(Stickel et al. \cite{sti04}), of which a small fraction does not have IRAS
detections. Most of the 1900 galaxies are spirals. The
measured 170\,${\mu}$m flux densities range from just below 0.5\,Jy up
to $\sim 600$\,Jy.

\subsubsection{2.2.1 The FIR spectral energy distribution: Dust temperatures, 
masses and luminosities}
\label{sssect:firsed}

The presence of a cold
dust emission component peaking longwards of 120\,${\mu}$m was
inferred from studies of the integrated SEDs of individual
galaxies (see Sect.~2.1.1), from statistical studies of
small samples (Kr\"ugel et al. \cite{krue98}; Siebenmorgen et
al. \cite{sie99}) and was confirmed and generalised by studies of the larger 
statistical samples mentioned above.
The latter studies also demonstrated the universality
of the cold dust component, showing it to be present within all types
of spirals (Tuffs \& Popescu \cite{tuf03}). 
The cold emission component predominantly arises from dust 
heated by the general diffuse interstellar medium and the warm component
from locally heated dust in HII regions, an interpretation consistent
with what has been seen in the ISOPHOT maps of nearby galaxies (see Sect.~2.1.1)
and with self-consistent modelling of the UV-FIR SEDs (see Sect.~2.3 and
Popescu \& Tuffs \cite{poptuf05}).

The cold dust component is most prominent 
in the most
``quiescent'' galaxies, like those contained in the IVCD sample, where
the cold dust temperatures were found to be broadly distributed, with
a median of 18\,K (Popescu et al. \cite{pop02}), some $8-10$\,K lower than
would have been predicted by IRAS. 
The corresponding dust masses were
correspondingly found to be increased by factors of typically $2-10$
(Stickel et al. \cite{sti00}) for the Serendipity Sample and by factors $6-13$
(Popescu et al. \cite{pop02}) for the IVCD sample, with respect to previous
IRAS determinations. As a consequence, the derived gas-to-dust ratios
are much closer to the canonical value of $\sim 160$ for the Milky Way
(Stickel et al. \cite{sti00}, Contursi et al. \cite{con01}; see also 
Haas et al. \cite{haa98a}
for M~31), but with a broad distribution of values (Popescu et
al. \cite{pop02}). 

It was found that the cold dust component provides not only
the bulk of the dust masses, but even the bulk of the FIR luminosity,
in particular for the case of the most quiescent spirals, like those
in the IVCD sample. In contrast to the SEDs found by the other ISOPHOT
studies, which typically peaked at around 170\,${\mu}$m, Bendo et
al. \cite{ben03} derived spatially integrated SEDs typically
peaking at around 100\,${\mu}$m. The result of Bendo et al.  may
reflect the fact that these observations were single pointings, made
towards the nucleus of resolved galaxies 
extending (in the main) beyond the field of view of ISOPHOT, and were therefore
biased towards nuclear emission, which is warmer than the extended
cold dust emission missed (or only partially covered) by these
measurements. Nevertheless, the measurements of Bendo et al. constitute a 
useful probe of the FIR emission of the inner disks. This emission 
(normalised to K band emission) was found to increase along the Hubble 
sequence (Bendo et al. \cite{ben02b}).

Since the FIR carries most of the dust luminosity, it is interesting to
  re-evaluate the question of the fraction of stellar photons converted via
  grains into IR photons, taking into account the comprehensive
  measurements of the cold dust emission component made available by
  ISOPHOT. This was done by
Popescu \& Tuffs \cite{poptuf02a}, who showed that the mean percentage of stellar
light reradiated by dust is $\sim30\%$ for the Virgo Cluster
spirals contained in the IVCD sample. This study also included the dust
  emission radiated in the NIR-MIR range. The fact that the mean value of 
$\sim30\%$ found for the Virgo Cluster spirals is the same as the canonical value
  obtained for the IRAS Bright Galaxy Sample (BGS; Soifer \& Neugebauer 
\cite{soi91})
  is at first sight strange, since IRAS was not sensitive to the cold dust
  component. However the BGS sample is an IR selected sample and 
  biased towards galaxies with higher dust luminosities, while the Virgo 
  sample is optically selected and contain a full representation of quiescent 
  systems. So the deficit in FIR emission caused by sample selection criteria
  for the Virgo sample is compensated for by the inclusion of the cold dust 
 component. 
Popescu \& Tuffs \cite{poptuf02a} also found evidence for an
increase of the ratio of the dust emission to the total stellar
emitted output along the Hubble sequence, ranging from typical values
of $\sim 15\%$ for early spirals to up to $\sim 50\%$ for some late
spirals. This trend was confirmed by Boselli et al. \cite{bos03a} who further
utilised the new ISO data on dust emission to constrain the
corresponding absorption of starlight and thus improve extinction
corrections (using the technique pioneered by Xu \&
Buat \cite{xb95} for the IRAS data).

\subsubsection{2.2.2 Two outstanding questions of the IRAS era}
\label{sssect:quest}

\medskip
\noindent
{\it The radio-FIR correlation}

One of the most surprising discoveries of the IRAS all-sky survey was
the very tight and universal correlation between the spatially
integrated FIR and radio continuum emissions (de Jong et al. \cite{dej85}; 
Helou et al. \cite{hel85}; Wunderlich et al. \cite{wun87}; see V\"olk \& Xu 
\cite{vx94} for a review). 
However all
the pre-ISO studies of the FIR/radio correlation were based on FIR
luminosities derived from the IRAS 60 and 100\,${\mu}$m flux
densities, and thus were missing the bulk of the cold dust
luminosity. The ISOPHOT measurements at 60, 100 and 170\,${\mu}$m were
used to redefined the FIR/radio correlation (Pierini et al. \cite{pie03}) 
for a 
statistical sample of spiral galaxies. The inclusion of
the cold dust component was found to produce a tendency for the total
FIR/radio correlation to become more non-linear than inferred from the 
IRAS
  60 and 100\,${\mu}$m\ observations. The use of the three FIR
wavelengths also meant that, for the first time, the correlation could
be directly derived for the warm and cold dust emission
components. 

The cold FIR/radio correlation was found to be slightly
non-linear, whereas the warm FIR/radio correlation is linear. Because
the effect of disk opacity in galaxies would introduce a non-linearity in
the cold-FIR/radio correlation, in the opposite sense to that
observed, it was argued that both the radio and the FIR emissions are
likely to have a non-linear dependence on SFR. For the radio emission an
enhancement of the small free-free component with SFR can account for this
effect. For the cold FIR emission a detailed analysis of the dependence
of local absorption and opacity of the diffuse medium on SFR is required to 
understand the non-linear trend of
the correlation (see Pierini et al. \cite{pie03}).

\begin{figure}[htb] 
\includegraphics[scale=0.40]{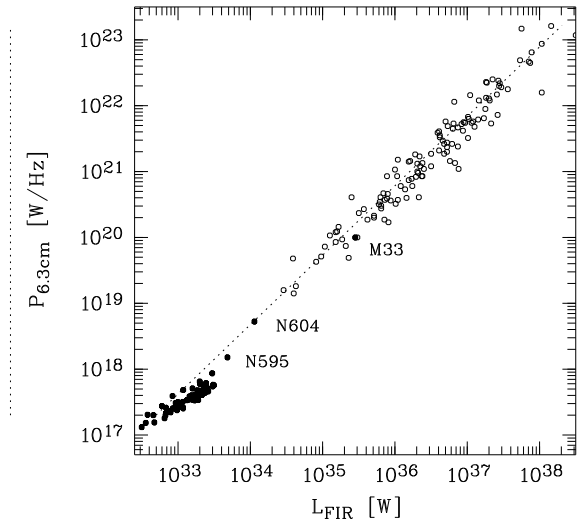}

\caption{Plot of the monochromatic radio luminosity versus the FIR
luminosities for M~33 (Hippelein et al. \cite{hip03}) and its star-forming
regions (filled circles) together with the data for the Effelsberg
100-m galaxy sample (Wunderlich et al. \cite{wun87}, open
circle). The dotted line has a slope of 1.10 (Wunderlich \& Klein
\cite{wk88}).}

\label{fig:f7}
\end{figure}

The improved angular resolution of ISO compared with IRAS also allowed a more 
detailed examination of the local FIR-radio correlation on sub-kpc size 
galactic substructures. Hippelein et al. \cite{hip03} established the 
correlation for the
star-forming regions in M~33. This correlation is shown in
Fig.~7, overplotted on the correlation for integrated
emission from galaxies. It is apparent that the local correlation has
a shallower slope (of the order of 0.9) than for the global
correlation. It was argued that the local correlation is attributable to the
increase with SFR of dust absorption in increased dust densities, and to local
synchrotron emission from within supernova remnants, still confining their 
accelerated electrons. Both emission components play only a minor role in the 
well known global radio-FIR correlation, that depends on the dominant 
large-scale absorption/re-emission properties of galaxies.

\medskip
\noindent
{\it Far-infrared emission as a star-formation tracer}

In the FIR, the correlation with the most widely-used star-formation
tracer, H${\alpha}$, is extremely non-linear, which has long led authors to
suspect that the FIR emission is the result of more than one component
(see e.g. Lonsdale-Persson \& Helou \cite{lon87}). With ISOPHOT the correlation was separately established for the warm and cold dust emission components.
A good linear correlation was found between the
warm FIR luminosities (normalised to the K band luminosity) derived for the 
IVCD sample and their H$\alpha$
EW (Popescu et al. \cite{pop02}). This is in agreement with the assumption
that the warm dust component is mainly associated with dust locally
heated within star-formation complexes. The scatter in the correlation was
  attributed to a small component of warm emission from the diffuse
disk (produced either by transiently heated grains or by grains heated by
the old stellar population), as well as to the likely variation in HII region
dust temperatures within and between galaxies.
A good but non-linear 
correlation was 
found between the cold FIR luminosities of the
galaxies from the IVCD sample and their H$\alpha$ EW, in the sense that 
FIR  increases more slowly than H$\alpha$. Since the bulk of the
cold FIR emission arises from the diffuse disk, the existence of this 
correlation implies that the grains in the diffuse disk are mainly powered by 
the UV photons (see also Sect.~2.3 and Popescu \& Tuffs \cite{poptuf05}). 
The non-linearity of the
correlation is consistent with there being a higher contribution from optical
photons in heating the grains in more quiescent galaxies.

For the late-type galaxies in the Coma and A1367 clusters, Contursi et
al. \cite{con01} derived the relationships between the IR flux
densities at 200, 170, 120, 100 and 60\,${\mu}$m, normalised to the H band flux,
as a function of the H${\alpha}$ EW. It was found that the poorer
correlation is in the 200\,${\mu}$m band and that the values of the
fitted slopes decrease as the FIR wavelength increases. These results
should be interpreted in terms of the increasing contribution of the
diffuse component with increasing FIR wavelength.

\begin{figure}[htb] 
\includegraphics[scale=0.45]{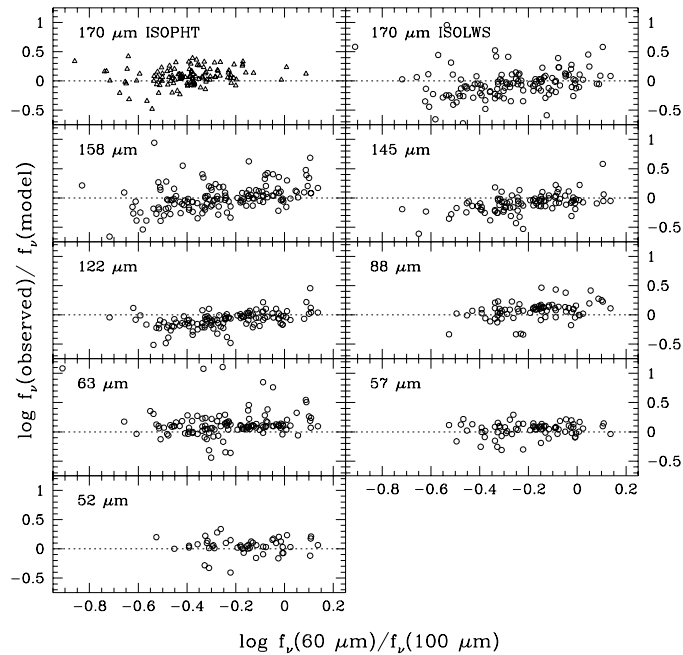}
\caption{Comparison of observed FIR continuum levels observed
by ISO with the model predictions of Dale \& Helou \cite{dalhel02}. The circles
derive from the ISO LWS templates and the triangles represent
170\,${\mu}$m data from the ISOPHOT Serendipity Survey (Stickel et
al. \cite{sti00}). }

\label{fig:f8}
\end{figure}

Using the ISOPHOT observations from Tuffs et al. \cite{tuf02a}, \cite{tuf02b} 
and Stickel et al. \cite{sti00} in combination with UV and K band photometry,
Pierini \& M\"oller \cite{piemoe03} have tried to quantify the effects of
optical heating and disk opacity on the derivation of SFR from FIR
luminosities.  For this they investigated trends in the ratio of the
far-IR luminosity to the intrinsic UV luminosity, L$_{dust}$/L$_{UV}$,
with both disk opacity and disk mass (as measured by the intrinsic
K-band luminosity). Using a separate relation between disk opacity and
K band luminosity they were able re-express L$_{dust}$/L$_{UV}$ in
terms of a single variable, the galaxy mass.
In this way they found evidence for the
relative importance of optical photons in heating dust to increase
with increasing galaxy mass.

\subsection{2.3 Quantitative interpretation of FIR SEDs}
\label{ssec:quant}

When considered in isolation, the FIR luminosity of normal galaxies is
a poor estimator of the SFR.  
There are two reasons for
this.  Firstly, these systems are only partially opaque to the UV
light from young stars, exhibiting large variations in the escape
probability of UV photons between different galaxies. Secondly, the
optical luminosity from old stellar populations can be so large in
comparison with the UV luminosity of young stellar populations that a
significant fraction of the FIR luminosity can be powered by optical
photons, despite the higher probability of absorption of UV photons
compared to optical photons.

\begin{figure}[htb] 
\includegraphics[scale=0.53]{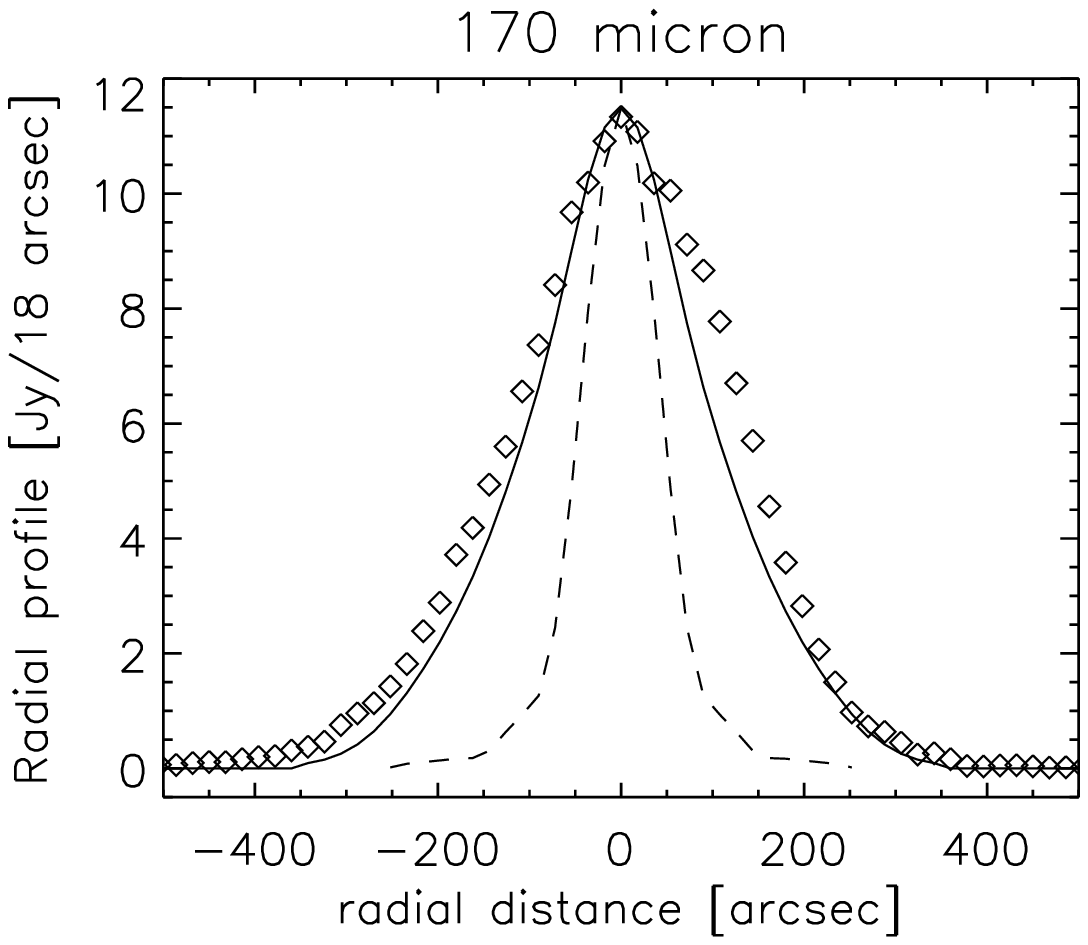}
\includegraphics[scale=0.40]{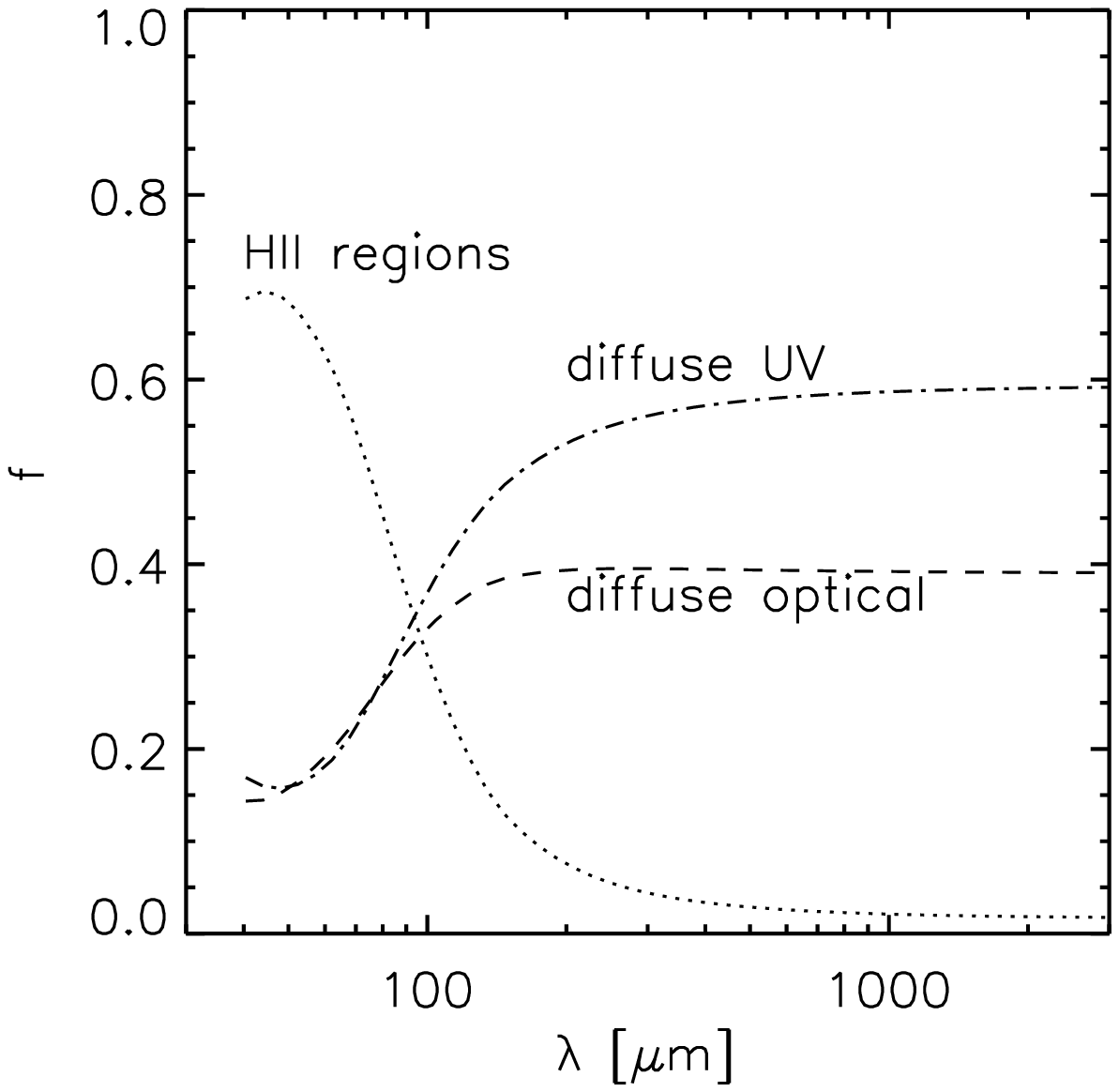}
\caption{Left: The radial profile of NGC~891 at 170\,${\mu}$m (Popescu et
al. \cite{pop04}) produced by integrated the emission parallel to the minor
axis of the galaxy for each bin along the major axis. Solid line:
model prediction; diamonds: observed profile; dotted line: beam
profile.
Right: The fractional contribution of the three stellar components to the FIR
emission of NGC~891 (Popescu et al. \cite{pop00b}).}

\label{fig:f9}
\end{figure}

One step towards a quantitative interpretation of FIR SEDs was
achieved by the semi-empirical models of Dale et al. \cite{dal01} and Dale \&
Helou \cite{dalhel02}. These authors have used ISO observations to develop a
family of templates to fit the variety of the observed forms of the IR
SEDs. This work assumes a power law distribution of dust masses with
local radiation field intensity to provide a wide range of dust
temperatures. It appears to indicate that the SEDs of star-forming
galaxies can be fitted by a one-parameter family of curves
(characterised by the 60/100\,${\mu}$m colour), determined essentially
by the exponent, $\alpha$, of the power law distribution of dust
masses with the radiation field intensity, where the radiation field
is assumed to have the colour of the local ISRF. This calibration
method has been extensively used to predict FIR flux densities
longwards of 100\,${\mu}$m for the galaxies not observed by ISOPHOT. A
comparison of the FIR continuum levels observed by ISO with
the Dale \& Helou \cite{dalhel02} model predictions is shown in
Fig.~8.  However, a quantitative interpretation of dust
emission in terms of SFRs and star-formation histories
requires a combined analysis of the UV-optical/FIR-submm SEDs,
embracing a self-consistent model for the propagation of the photons.

There are a few such models in use which incorporate various
geometries for the stellar populations and dust (Silva et al. \cite{sil98},
Bianchi et al. \cite{bia00}, Charlot \& Fall \cite{chafal00}, Popescu et
al. \cite{pop00b}; see review by Popescu \& Tuffs \cite{poptuf05}).  We will 
concentrate 
here on the model of Popescu et al. \cite{pop00b}, since this is the only model which has been used to make direct
predictions for the spatial distribution of the FIR emission for
comparison with the ISOPHOT images. This model also ensures that the
geometry of the dust and stellar populations is consistent with
optical images. Full details are given by Popescu et al. \cite{pop00b},
Misiriotis et al. \cite{mis01}, Popescu \& Tuffs \cite{poptuf02b}
 and Tuffs et al. \cite{tuf04}. 
In brief, the model includes solving
the radiative-transfer problem for a realistic distribution of
absorbers and emitters, considering realistic models for dust, taking
into account the grain-size distribution and stochastic heating of
small grains and the contribution of HII regions.  The FIR-submm SED
is fully determined by just three parameters: the star-formation rate
$SFR$, the dust mass $M_{\rm dust}$ associated with the young stellar
population, and a factor $F$, defined as the fraction of non-ionising
UV photons which are locally absorbed in HII regions around the
massive stars. A self-consistent theoretical approach to the calculation of the
$F$ factor is given by Dopita et al. \cite{dop05} in the context of modelling SEDs
of starburst galaxies.

Popescu et al. \cite{pop00b} illustrated their model with the example of the 
edge-on galaxy NGC~891, which has been extensively observed at all 
wavelengths (including the complete submm range; Dupac et al. \cite{dup03}), 
and also mapped with ISOPHOT at 170 and 200\,${\mu}$m (Popescu et
al. \cite{pop04}). 
A particularly stringent test of the model was to compare its
prediction for the radial profile of the diffuse dust emission
component near the peak of the FIR SED with the observed radial
profiles. This comparison (Popescu et al. \cite{pop04}; see also Fig.~9, left 
panel) showed a remarkable agreement. 
The excess emission in the observed profile with respect to
the predicted one for the diffuse emission was explained in terms of two 
localised sources. 

At 170 and 200\,${\mu}$m the model for NGC~891 predicts that the bulk
of the FIR dust emission is from the diffuse component. The close
agreement between the data and the model predictions, both in
integrated flux densities, but especially in terms of the spatial
distribution, constitutes a strong evidence that the large-scale
distribution of stellar emissivity and dust predicted by the model is
in fact a good representation of NGC~891. In turn, this supports the
prediction of the model that the dust emission in NGC~891 is
predominantly powered by UV photons.

Depending on the FIR/submm wavelength, the UV-powered dust emission
arises in different proportions from within the localised component (HII
regions) and from the diffuse component (Fig.~9; right
panel).  For example at 60\,${\mu}$m, 61$\%$ of the FIR emission in NGC~891 was
predicted to be powered by UV photons locally absorbed in star-forming
complexes, 19$\%$ by diffuse UV photons in the weak radiation fields
in the outer disk (where stochastic emission predominates), and 20$\%$
by diffuse optical photons in high energy densities in the inner part
of the disk and bulge. At 100\,${\mu}$m the prediction was that there
are approximately equal contributions from the diffuse UV, diffuse
optical and locally absorbed UV photons. At 170, 200\,${\mu}$m and
submm wavelengths, most of the dust emission in NGC~891 was predicted
to be powered by the diffuse UV photons. The analysis described above
does not support the preconception that the weakly heated cold dust
(including the dust emitting near the peak of the SED sampled by the
ISOPHOT measurements presented here) should be predominantly powered
by optical rather than UV photons. The reason is as follows: the
coldest grains are those which are in weaker radiation fields, either
in the outer optically thin regions of the disk, or because they are
shielded from radiation by optical depth effects. In the first
situation the absorption probabilities of photons are controlled by
the optical properties of the grains, so the UV photons will dominate
the heating. The second situation arises for dust associated with the
young stellar population, where the UV emissivity far exceeds the
optical emissivity.

\section{3. Dwarf Galaxies}
\subsection{3.1 Cold dust surrounding dwarf galaxies}
\label{ssec:surround}

Gas-rich dwarf galaxies and in particular Blue Compact Dwarfs (BCDs)
were originally expected to have their FIR emission dominated by dust
heated locally in HII regions. Temperatures of 30\,K or more were
anticipated. This was the a priori expectation in particular for the
BCDs, and became the standard interpretation for the IRAS results
obtained for these systems.  Hoffman et al. \cite{hof89}, Helou et
al. \cite{hel88} and Melisse \& Israel \cite{melisr94} each found that the
60/100\,${\mu}$m colours of BCDs were clearly warmer than those of
spirals.

The IVCD Survey (Tuffs et al. \cite{tuf02a}, \cite{tuf02b}) changed that 
simple picture of
the FIR emission from dwarf galaxies.  The IVCDS included measurements
at 60, 100, and (for the first time) at 170\,${\mu}$m of 25 optically
selected gas-rich dwarf galaxies. The observations at 60 and
100\,${\mu}$m were consistent with the previous IRAS results, though extending knowledge of these systems to
lower luminosities. Unexpectedly however, high ratios of
170/100\,${\mu}$m\ luminosities were found in many of the surveyed
systems. Such long-wavelength excesses were found both in relatively
high-luminosity dwarfs, such as VCC655, as well as in fainter objects
near the limiting sensitivity of the survey. These observations imply
the presence of large amounts of cold dust. 

As shown by Popescu et al. \cite{pop02}, it seems unlikely that the cold
dust resides in the optically thick molecular component associated
with star-formation regions, since the implied dust masses
would be up to an order of magnitude greater than those typically
found in giant spirals.  Such large masses could not have been
produced through star-formation within the dwarfs over their
lifetime. One alternative is that the dust
originates and still resides outside the optical extent of these
galaxies. In fact, some evidence for this was provided by the ISOPHOT
observations themselves, since they were made in the form of
scan maps, from which estimates of source sizes could be
determined. Even with the relatively coarse beam (1.6$^{\prime}$ FWHM at
170\,${\mu}$m), the extended nature of the sources could be clearly
seen in a few cases 
for which the FWHM for the 170\,${\mu}$m emission exceeds the
optical diameters of the galaxies (to 25.5Bmag/arcsec$^2$) by factors
of between 1.5 and 3.5. It is interesting to note that two of these
galaxies (VCC~848 and VCC~81) have also been mapped in HI (Hoffman et
al. \cite{hof96}), revealing neutral hydrogen sizes comparable to the
170\,${\mu}$m extent. This raises the possibility that the
cold dust is embedded in the extended HI gas, external to the optical
galaxy.  This would be analogous to the case of the edge-on spiral
NGC~891, where Popescu \& Tuffs \cite{poptuf03} discovered a cold-dust
counterpart to the extended HI disk (see Sect.~2.1.2). In
this context the main observational difference between the giant
spiral and the dwarfs may be that for the dwarfs the integrated
170\,${\mu}$m emission is dominated by the extended emission component
external to the main optical body of the galaxy, whereas for the giant
spirals the long-wavelength emission predominantly arises from within
the confines of the optical disk of the galaxy. 

Apart from the SMC (which is discussed in Sect.~3.2 and is
too extended for ISOPHOT to map beyond its optical
extent), only three dwarf galaxies were observed by ISOPHOT in the
field environment (all Serendipity Survey sources; Stickel et
al. \cite{sti00}). All three sources have comparable flux densities at 100 and
170\,${\mu}$m. But the small statistics mean that it is still an open
question to which extent the cold dust emission associated with the
extended HI component in dwarf galaxies is a cluster phenomenon or
not. Of potential relevance to the
origin of the dust seen surrounding Virgo dwarfs is the discovery by ISOPHOT 
of the source ``M~86-FIR'' in the halo of M~86 (Stickel et al. 
\cite{sti03}) which has no optical counterpart. It may be a ``relic'' ISM
stripped from a spiral galaxy in the Virgo Cluster. 

The existence of large quantities of dust surrounding gas-rich dwarf
galaxies may have important implications for our understanding of the
distant Universe.  According to the hierarchical galaxy formation
scenarios, gas-rich dwarf galaxies should prevail at the earliest
epochs. We would then expect these same galaxies to make a higher
contribution to the total FIR output in the early Universe, certainly
more than previously expected.

\subsection{3.2 Infrared emission from within dwarf galaxies}
\label{ssec:smc}

The distribution of cold dust within dwarf galaxies could be studied
in only one case, namely in the resolved (1.5$^{\prime}$ resolution)
170\,${\mu}$m ISOPHOT map of the Small Magellanic Cloud (Wilke et
al. \cite{wil03}, \cite{wil04}).  The 170\,${\mu}$m ISO map of the SMC 
reveals a wealth of structure, not only
consisting of filamentary FIR emitting regions, but also of numerous
(243 in total) bright sources which trace the bar along its major axis
as well as the bridge which connects the SMC to the LMC. Most of the
brighter sources have cold components, associated with molecular
clouds. The discrete sources were found to contribute $28\%$, $29\%$
and $36\%$ to the integrated flux densities at 60, 100 and
170\,${\mu}$m, respectively. 
The SED was modelled by the superposition of 45\,K, 20.5\,K and 10\,K
blackbody components with emissivity index $\beta$=2. The average dust colour 
temperature (averaged over all pixels of the 170/100 colour map) was found to 
be $T_{\rm D}=20.3\,$K.

Bot et al. \cite{bot04} have compared the FIR map of the SMC with
an HI map of similar resolution. This reveals a good spatial
correlation of the two emissions in the diffuse regions of the maps
(regions that fall outside of the correlation are either hot star-forming 
regions, or cold molecular clouds with no associated
HI). Adding the IRAS data allows them to compute the FIR emissivity
per unit H atom. Bot et al. \cite{bot04} found that this emissivity is lower
than in the Milky Way, and in fact it is even lower than the lower
metallicity ($Z_{\odot}/10$) of the SMC would imply, suggesting that
depletion mechanisms at work in the ISM have more than a linear
dependence on metallicity.

Although FIR emission from dwarf galaxies has been associated only with 
gas-rich
dwarfs, in one particular case such emission has been detected in a dwarf
elliptical galaxy, as well. This is the case of NGC~205, one of the companions
of M~31, classified as a peculiar dE5. This galaxy shows signatures of recent
star-formation (Hodge \cite{hod73}) and of extended HI emission 
(Young \& Lo \cite{yl97}),
and was detected by IRAS (Rice et al. \cite{ric88}, Knapp et al. \cite{kna89}, 
with a SED 
steeply rising between 60 and 100\,${\mu}$m. Based on  
ISOPHOT observations, Haas \cite{haa98b} showed that the FIR emission is
resolved and similar to that seen in HI. He also presented evidence for 
a very cold dust component, of 10 K, coming from the center of the galaxy.


Galliano et al. \cite{gal03}  modelled the UV-optical/MIR-FIR-submm SED of 
the low metallicity nearby dwarf galaxy NGC1569.
(Fig.~10). This study is
noteworthy in that it constrains the grain size distribution through
the MIR-FIR ISOCAM and ISOPHOT observations and therefore gives 
more specific information about grain properties in dwarf galaxies. 
The results indicate a paucity of Polycyclic Aromatic Hydrocarbons (PAHs) due 
to an enhanced destruction in the intense ambient UV radiation field, as well 
as an overabundance
(compared to Milky Way type dust) of small grains of size $\sim
3$\,nm, possibly indicative of a redistribution of grain sizes through
the effect of shocks.

\begin{figure}[htb] 
\includegraphics[scale=0.5]{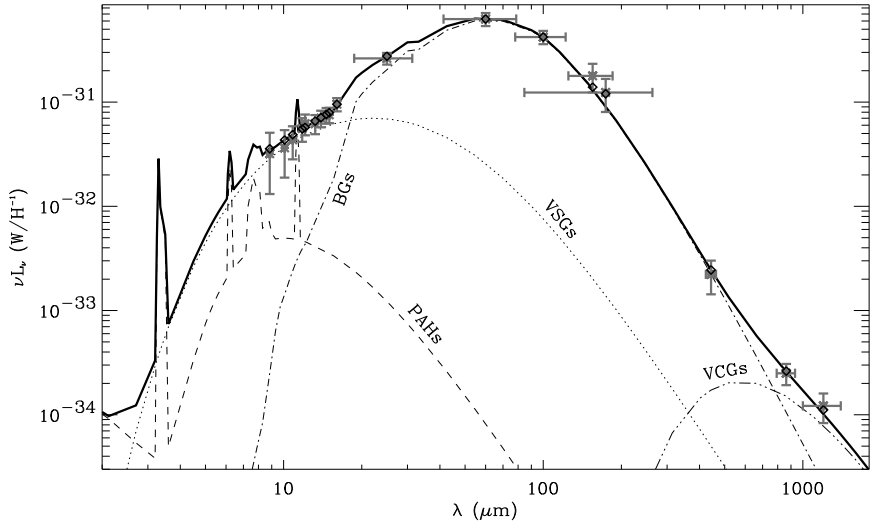}

\caption{NGC~1569 observations and modeled SED from Galliano et
al. \cite{gal03}. The data are indicated by crosses: vertical bars are the
errors on the flux density values and the horizontal bars indicate the
widths of the broadbands. The lines show the predictions of the dust
model with its different components. Diamonds indicate the model
predictions integrated over the observational broadbands and
colour-corrected.}

\label{fig:f10}
\end{figure} 

\section{4. Conclusions and Outlook}

ISOPHOT has not only advanced the knowledge of IR properties of normal
galaxies but has also made unexpected discoveries. 
The bulk of the emission from dust has been measured, 
revealing cold dust in copious quantities. This dust is present in all types 
of normal galaxies and is predominantly distributed in a diffuse disk with an 
intrinsic scalelength exceeding that of the stars. Cold 
dust has been found beyond the optical regions of isolated
galaxies, associated with the extended HI disks of spiral galaxies or with the
HI envelopes of dwarf galaxies. The fraction of the
bolometric luminosity radiated by dust has been measured for the first time. 
Realistic 
geometries for stars and dust have been derived from ISOPHOT imaging 
observations, enabling the contribution of the various stellar populations to 
the dust heating to be accurately derived. 
This enormous advancement in the understanding of normal galaxies in the 
nearby Universe has laid the foundation for more detailed 
investigations with Spitzer and Herschel.
A clear priority
is to increase the number of galaxies with detailed imaging information and to
provide better statistics on carefully selected samples, especially those
selected in the optical/NIR bands. Ultimately, the improved sensitivity of 
the new
infrared space observatories will allow knowledge of the dust emission from
normal galaxies to be extended beyond the nearby universe.

\begin{theacknowledgments}
The authors would like to take this opportunity to thank all the individuals
that helped make the ISO mission a success.  R.J. Tuffs and C.C. Popescu 
would also like to thank Heinrich J. V\"olk for enlightening discussions.
\end{theacknowledgments}

\IfFileExists{\jobname.bbl}{}
 {\typeout{}
  \typeout{******************************************}
  \typeout{** Please run "bibtex \jobname" to optain}
  \typeout{** the bibliography and then re-run LaTeX}
  \typeout{** twice to fix the references!}
  \typeout{******************************************}
  \typeout{}
 }

\begin{thebibliography}{99}
\bibitem[\protect\citeauthoryear{Alton et al.}{1998b}]{alt98b} 
Alton P.B., Bianchi S., Rand R.J., Xilouris, E.M., Davies, J.I. 
et al., 1998b, ApJ, 507, L125

\bibitem[\protect\citeauthoryear{Alton et al.}{1998a}]{alt98a} 
Alton, P.B., Trewhella, M., Davies, J.I., Evans, R., Bianchi, S., et
al. 1998a, A\&A, 335, 807


\bibitem[\protect\citeauthoryear{Bendo et al.}{2002a}]{ben02a} 
Bendo, G.J., Joseph, R.D., Wells, M., Gallais, P., Haas, M. et 
al. 2002a, AJ 123, 3067

\bibitem[\protect\citeauthoryear{Bendo et al.}{2002b}]{ben02b} 
Bendo, G.J., Joseph, R.D., Wells, M., Gallais, P., Haas, M. 
et al. 2002b, AJ 124, 1380

\bibitem[\protect\citeauthoryear{Bendo et al.}{2003}]{ben03} 
Bendo, G.J., Joseph, R.D., Wells, M., Gallais, P., Haas, M. 
et al. 2003, AJ 125, 2361


\bibitem[\protect\citeauthoryear{Bianchi et al.}{2000}]{bia00} 
Bianchi, S., Davies, J. I., Alton, P. B. 2000, A\&A 359, 65


\bibitem[\protect\citeauthoryear{Bingelli et al.}{1993}]{bin93} 
Binggeli, B., Popescu, C.C. \& Tammann, G.A. 1993, A\&AS, 
98, 275

\bibitem[\protect\citeauthoryear{Bingelli et al.}{1985}]{bin85} 
Binggeli, B., Sandage, A. \& Tammann, G.A. 1985, AJ, 90, 1681

\bibitem[\protect\citeauthoryear{Bot et al.}{2004}]{bot04}
Bot, C., Boulanger, F., Lagache, G., Cambr\'esy, L., Egret, D. 2004
A\&A 423, 567

\bibitem[\protect\citeauthoryear{Boselli et al.}{2002}]{bos02} 
Boselli, A., Gavazzi, G., Lequeux, J.  Pierini, D. 2002, A\&A
  385, 454

\bibitem[\protect\citeauthoryear{Boselli et al.}{2003a}]{bos03a} 
Boselli, A., Gavazzi, G., Sanvito, G. 2003a, A\&A 402, 37

\bibitem[\protect\citeauthoryear{Boselli et al.}{1997}]{bos97} 
Boselli, A., Lequeux, J., Contursi, A. Gavazzi, G., Boulade, O., et al.
1997, A\&A, 324, L13

\bibitem[\protect\citeauthoryear{Boselli et al.}{2003b}]{bos03b} 
Boselli, A., Sauvage, M., Lequeux, J., Donati, A, \& Gavazzi,
G. 2003b, A\&A 406, 867




\bibitem[\protect\citeauthoryear{Charlot \& Fall}{2000}]{chafal00} Charlot, S., \& Fall, S.M. 2000, ApJ 539, 718


\bibitem[\protect\citeauthoryear{Contursi et al.}{2001}]{con01}
Contursi, A., Boselli, A., Gavazzi, G., Bertagna, E., Tuffs, R.,
Lequeux, J. 2001, A\&A 365, 11

\bibitem[\protect\citeauthoryear{Contursi et al.}{1998}]{con98}
Contursi, A., Lequeux, J., Hanus, M., Heydari-Malayeri, M., Bonoli,
C., Bosma, A., et al. 1998, A\&A 336, 662

\bibitem[\protect\citeauthoryear{Contursi et al.}{2000}]{con00}
Contursi, A., Lequeux, J., Cesarsky, D., Boulanger, F., Rubio, M.,
Hanus, M. et al. 2000, A\&A 362, 310

  

\bibitem[\protect\citeauthoryear{Dale \& Helou}{2002}]{dalhel02} Dale, D. \& Helou, G. 2002, ApJ 576, 159

\bibitem[\protect\citeauthoryear{Dale et al.}{2001}]{dal01} 
Dale, D. A., Helou, G., Contursi, A., Silbermann, N. A. \& Kolhatkar,
S. 2001, ApJ, 549, 215

\bibitem[\protect\citeauthoryear{Dale et al.}{1999}]{dal99} 
Dale, D. A., Helou, G., Silbermann, N. A., Contursi, A., Malhotra, S.,
Rubin, R. H. 1999, AJ 118, 2055
  
\bibitem[\protect\citeauthoryear{Dale et al.}{2000}]{dal00}
Dale, D. A., Silbermann, N. A., Helou, G., Valjavec, E., Malhotra, S.,
Beichmann, C. A., et al. 2000, AJ 120, 583

\bibitem[\protect\citeauthoryear{Davies et al.}{1998}]{dav98} 
Davies, J. I., Alton, P., Bianchi, S. \& Trewhella, M. 1998, MNRAS
300, 1006

\bibitem[\protect\citeauthoryear{Davies et al.}{1999}]{dav99} Davies, J.I., Alton, P., Trewhella, M., Evans, R., \& 
Bianchi, S. 1999, MNRAS, 304, 495


\bibitem[\protect\citeauthoryear{de Jong et al.}{1984}]{dej84} de Jong, T., Clegg, P.E., Soifer, B.T.., et al. 1984, ApJ, 278, 
L67
 
\bibitem[\protect\citeauthoryear{de Jong et al.}{1985}]{dej85} de Jong, T., Klein, U., Wielebinski, R., Wunderlich,
E. 1985, A\&A, 147, L6


\bibitem[\protect\citeauthoryear{Dopita et al.}{2005}]{dop05} Dopita, M.A., Groves, B.A., Fischera, J., Sutherland, R.S., Tuffs,
  R.J. et al. 2005, ApJ, in press

\bibitem[\protect\citeauthoryear{Dupac et al.}{2003}]{dup03} Dupac, X., del Burgo, C., Bernard, J.-P., et al. 2003, MNRAS 344,
  105

\bibitem[\protect\citeauthoryear{Ferguson et al.}{1998}]{ferg98} Ferguson, A., Gallagher, J.S. \& Wyse, R. 1998, AJ 116, 673

\bibitem[\protect\citeauthoryear{Ferrara et al.}{1991}]{fer91} 
Ferrara, A., Ferrini, F., Barsella, B., \& Franco, J. 1991, 
ApJ 381, 137




\bibitem[\protect\citeauthoryear{Galliano et al.}{2003}]{gal03} 
Galliano, F., Madden, S.C., Jones, A.P., Wilson, C.D., Bernard, J.-P.,
et al. 2003, A\&A 407, 159

\bibitem[\protect\citeauthoryear{Gerhard \& Silk}{1996}]{gs96} Gerhard, O. \& Silk, J. 1996, ApJ 472, 34


\bibitem[\protect\citeauthoryear{Haas}{1998}]{haa98b} Haas, M. 1998, 
A\&A 337, L1

\bibitem[\protect\citeauthoryear{Haas et al.}{2002}]{haa02}
Haas, M., Klaas, U., Bianchi, S. 2002, A\&A 385, L23

\bibitem[\protect\citeauthoryear{Haas et al.}{1998}]{haa98a} 
Haas, M., Lemke, D., Stickel, M.,  Hippelein, H., Kunkel, M. 
et al. 1998, A\&A, 338, L33

\bibitem[\protect\citeauthoryear{Haas et al.}{2004}]{haa04} 
Haas, M., Siebenmorgen, R., Leipski, C., Ott, S., Cunow, B.,
Meusinger, H., et al. 2004, A\&A 419, L49

\bibitem[\protect\citeauthoryear{Helou}{1986}]{hel86}
Helou, G. 1986, ApJ 311, 33

\bibitem[\protect\citeauthoryear{Helou et al.}{1988}]{hel88}
Helou, G., Khan, I. R., Malek, L., \& Boehmer, L. 1988, ApJS, 
68, 151

\bibitem[\protect\citeauthoryear{Helou et al.}{1996}]{hel96} 
Helou, G., Malhotra, S., Beichman, C. A., Dinerstein, H., Hollenbach,
D. J., Hunter, D. A., et al. 1996, A\&A 315, L157

\bibitem[\protect\citeauthoryear{Helou et al.}{2001}]{hel01}
Helou, G., Malhotra, S., Hollenbach, D. J., Dale, D. A., Contursi,
A. 2001, ApJ 548, L73

\bibitem[\protect\citeauthoryear{Helou et al.}{2004}]{hel04}
Helou, G., Roussel, H., Appleton, P., Frayer, D., Stolovy, S., et
al. 2004, ApJS, in press

\bibitem[\protect\citeauthoryear{Helou et al.}{1985}]{hel85} 
Helou, G., Soifer, B. T., \& Rowan-Robinson, M. R. 1985, ApJ, 
298, L7


\bibitem[\protect\citeauthoryear{Hippelein et al.}{2003}]{hip03} Hippelein, H., Haas, M., Tuffs, R.J., Lemke, D., Stickel,
M. et al. 2003, A\&A 407, 137

\bibitem[\protect\citeauthoryear{Hodge}{1973}]{hod73} Hodge, P. 1973, ApJ 182, 671
   
\bibitem[\protect\citeauthoryear{Hoffman et al.}{1989}]{hof89} Hoffman, G. L., Helou, G., Salpeter, E. E., Lewis, B. M. 
1989, ApJ, 339, 812

\bibitem[\protect\citeauthoryear{Hoffman et al.}{1996}]{hof96}
Hoffman, G. L., Salpeter, E. E., Farhat, B., Roos, T., Williams, H.,
and Helou, G. 1996, ApJS 105, 269

\bibitem[\protect\citeauthoryear{Hunter et al.}{2001}]{hun01} 
Hunter, D. A., Kaufman, M., Hollenbach, D. J., Rubin, R. H., Malhotra,
S., et al. 2001, ApJ 553, 121


\bibitem[\protect\citeauthoryear{Kessler et al.}{1996}]{kes96}
Kessler, M. F., Steinz, J. A., Anderegg, M. E., Clavel, J., Drechsel, G., 
et al. 1996, A\&A, 315, 27

\bibitem[\protect\citeauthoryear{Knapp et al.}{1989}]{kna89} Knapp, G.R., Guhathakurta, P., Kin, D.W., Jura, M. 1989, ApJS 70,
  329 



\bibitem[\protect\citeauthoryear{Kr\"ugel et al.}{1998}]{krue98} Kr\"ugel, E., Siebenmorgen, R., Zota, V., Chini, R. 1998, 
A\&A, 331, L9 

\bibitem[\protect\citeauthoryear{Leech et al.}{1999}]{lee99} 
Leech, K.J., V\"olk, H.J., Heinrichsen, I., Hippelein, H., Metcalfe,
L., et al. 1999, MNRAS, 310, 317

\bibitem[\protect\citeauthoryear{Lemke et al.}{1996}]{lem96}
Lemke, D., Klaas, U., Abolins, J.,
 \'{A}br\'{a}ham, P., Acosta-Pulido, J., et al. 1996, A\&A, 315, L64

\bibitem[\protect\citeauthoryear{Li \& Draine}{2002}]{li02}
Li, A., Draine, B. T. 2002, ApJ 572, 232

\bibitem[\protect\citeauthoryear{Lonsdale-Persson \& Helou}{1987}]{lon87}
Lonsdale-Persson C. J., Helou, G. 1987, ApJ 314, 513


\bibitem[\protect\citeauthoryear{Lu et al.}{2003}]{lu03} 
Lu, N., Helou, G., Werner, M. W., Dinerstein, H. L., Dale, D. A.,
et al. 2003, ApJ 588, 199


  

 


\bibitem[\protect\citeauthoryear{Malhotra et al.}{2001}]{mal01} 
Malhotra, S., Kaufman, M.J., Hollenbach, D., Helou, G., Rubin, R.H. et
al. 2001, ApJ 561, 766

\bibitem[\protect\citeauthoryear{Martin et al.}{2005}]{mar05} Martin, C. et al. 2005, ApJ Letters, in press

\bibitem[\protect\citeauthoryear{Melisse \& Israel}{1994}]{melisr94} Melisse, J. P. M., Israel, F. P. 1994, A\&A, 285, 51

\bibitem[\protect\citeauthoryear{Meijerink et al.}{2004}]{mei04} Meijerink, R.,
   Tilanus, R.P.J., Dullemond, C.P., Israel, F.P., \&  van der Werf, P.P. 2004,
   A\&A in press

\bibitem[\protect\citeauthoryear{Misiriotis et al.}{2001}]{mis01} Misiriotis A., Popescu, C.C., Tuffs, R.J., \& Kylafis, N.D. 
2001, A\&A, 372, 775





\bibitem[\protect\citeauthoryear{Pierini et al.}{2003a}]{pie03a}
Pierini, D., Leech, K. J., V\"olk, H. J.  2003a, A\&A, 397, 871


\bibitem[\protect\citeauthoryear{Pierini \& M\"oller}{2003}]{piemoe03} Pierini, D. \& M\"oller, C. 2003, MNRAS 346, 818 

\bibitem[\protect\citeauthoryear{Pierini et al.}{2003b}]{pie03} 
Pierini, D., Popescu, C. C., Tuffs, R. J., V\"olk, H. J. 2003b,
A\&A 409, 907

\bibitem[\protect\citeauthoryear{Pfenniger \& Combes}{1994}]{pc94} 
Pfenniger, D., Combes, F. 1994, A\&A 285, 94
\bibitem[\protect\citeauthoryear{Pfenniger et al.}{1994}]{pfe94} 
Pfenniger, D., Combes, F., Martinet, L. 1994, A\&A 285, 79

\bibitem[\protect\citeauthoryear{Popescu et al.}{2000b}]{pop00b} Popescu, C.C., Misiriotis A., Kylafis, N.D., Tuffs, R.J., 
\& Fischera, J., 2000b, A\&A, 362, 138

\bibitem[\protect\citeauthoryear{Popescu \& Tuffs}{2002a}]{poptuf02a} Popescu, C.C., Tuffs, R.J. 2002a, MNRAS 335, L41

\bibitem[\protect\citeauthoryear{Popescu \& Tuffs}{2002b}]{poptuf02b} 
Popescu, C.C., Tuffs, R.J. 2002b, Reviews in Modern Astronomy, vol 15., 
Edited by Reinhard E. Schielicke. Wiley, ISBN 352640404X, p.239

\bibitem[\protect\citeauthoryear{Popescu \& Tuffs}{2003}]{poptuf03} 
Popescu, C.C, Tuffs, R.J. 2003, A\&A 410, L21

\bibitem[\protect\citeauthoryear{Popescu \& Tuffs}{2005}]{poptuf05} 
Popescu, C.C, Tuffs, R.J. 2005, in "The Spectral Energy Distribution of 
Gas-Rich Galaxies: Confronting Models with Data", Heidelberg, 4-8 Oct. 2004, eds. C.C. Popescu \& R.J. Tuffs, AIP Conf. Ser., in press

\bibitem[\protect\citeauthoryear{Popescu et al.}{2000a}]{pop00a} 
Popescu, C.C., Tuffs, R.J., Fischera, J., V\"olk, H.J. 
2000a, A\&A, 354, 480

\bibitem[\protect\citeauthoryear{Popescu et al.}{2004}]{pop04} 
Popescu, C. C., Tuffs, R. J., Kylafis, N. D. \& Madore, B. F. 2004,
A\&A 414, 45

\bibitem[\protect\citeauthoryear{Popescu et al.}{2005}]{pop05} Popescu, C.C., Tuffs, R.J., Madore, B.F., Gil de Paz, A.,
V\"olk, H. et al. 2005, ApJ 619, L75
  
\bibitem[\protect\citeauthoryear{Popescu et al.}{2002}]{pop02} Popescu, C.C., Tuffs, R.J., V\"olk, H.J., Pierini, D.
\& Madore, B.F., 2002, ApJ 567, 221



\bibitem[\protect\citeauthoryear{Rice et al.}{1988}]{ric88} Rice, W., Lonsdale, C.J., \& Soifer, B.T., et al. 1988, ApJS 68,
  91 


  
\bibitem[\protect\citeauthoryear{Roussel et al.}{2001c}]{rouatl}
Roussel, H., Vigroux, L., Bosma, A., Sauvage, M., Bonoli, C., Gallais,
P., et al. 2001c, A\&A 369, 473



\bibitem[\protect\citeauthoryear{Schmitobreick et al.}{2000}]{smi00} 
Schmidtobreick, L., Haas, M., \& Lemke, D. 2000, A\&A 363, 917

\bibitem[\protect\citeauthoryear{Siebenmorgen et al.}{1999}]{sie99} Siebenmorgen, R., Kr\"ugel, E., Chini, R. 1999, A\&A, 351, 
495

\bibitem[\protect\citeauthoryear{Silva et al.}{1998}]{sil98} 
Silva, L., Granato, G. L., Bressan, A., Danese, L. 1998, ApJ 509, 103


\bibitem[\protect\citeauthoryear{Soifer \& Neugebauer}{1991}]{soi91} Soifer, B.T., \& Neugebauer, G. 1991, AJ, 101, 354


\bibitem[\protect\citeauthoryear{Stickel et al.}{2003}]{sti03} Stickel, M., 
Bregman, J.N., Fabian, A.C., White, D.A., \&
  Elmegreen, D.M. 2003, A\&A 397, 503

\bibitem[\protect\citeauthoryear{Stickel et al.}{2000}]{sti00} Stickel, M., Lemke, D., Klaas, U., Beichman, C.A., Rowan-Robinson,
  M. et al. 2000, A\&A 359, 865

\bibitem[\protect\citeauthoryear{Stickel et al.}{2004}]{sti04} Stickel, M., Lemke, D., Klaas, U., Krause, O., \& Egner,
  S. 2004, A\&A 422, 39


\bibitem[\protect\citeauthoryear{Swaters et al.}{1997}]{swa97} 
Swaters, R. A., Sancisi, R. \& van der Hulst, J. M. 1997, ApJ 491, 140


\bibitem[\protect\citeauthoryear{Trewhella et al.}{2000}]{tre00}
Trewhella, M., Davies, J. I., Alton, P. B., Bianchi, S., \& Madore,
B. F. 2000, ApJ 543, 153


\bibitem[\protect\citeauthoryear{Tuffs et al. }{1996}]{tuf96} Tuffs, R.J., Lemke, D., Xu, C. et al. 1996, A\&A 315, L149

\bibitem[\protect\citeauthoryear{Tuffs \& Gabriel}{2000}]{tg03} Tuffs, R.J. \& Gabriel, C., 2003, A\&A 410, 1075

\bibitem[\protect\citeauthoryear{Tuffs \& Popescu}{2003}]{tuf03} Tuffs, R.J. \& Popescu, C.C. 2003, 
in ``Exploiting the ISO Data Archive. Infrared Astronomy in the Internet Age'',
Siguenza, Spain 24-27 June, 2002. Eds. C. Gry et al., ESA SP-511, p. 239.
 
\bibitem[\protect\citeauthoryear{Tuffs et al. }{2002a}]{tuf02a} Tuffs, R.J., Popescu, C.C., Pierini, D., V\"olk, H. J., 
Hippelein, H. et al. 2002a, ApJS 139, 37

\bibitem[\protect\citeauthoryear{Tuffs et al. }{2002b}]{tuf02b} Tuffs, R.J., Popescu, C.C., Pierini, D., V\"olk, H. J., 
Hippelein, H. et al. 2002b, ApJS 140, 609

\bibitem[\protect\citeauthoryear{Tuffs et al. }{2004}]{tuf04} Tuffs, R.J., Popescu, C. C., V\"olk, H.J., Kylafis, N.D., \& 
Dopita, M.A. 2004, A\&A 419, 821


\bibitem[\protect\citeauthoryear{Valentijn \& van der Werf}{1999a}]{val99a} 
Valentijn, E. A., van der Werf, P. P. 1999a, in Proceedings of the
Conference ``The Universe as seem by ISO'', Eds. P. Cox \&
M. F. Kessler. ESA-SP 427., p.821

\bibitem[\protect\citeauthoryear{Valentijn \& van der Werf}{1999b}]{val99b}
Valentijn, E. A. \& van der Werf, P. P. 1999b, ApJ 522, L29

\bibitem[\protect\citeauthoryear{Valentijn et al.}{1996}]{val96}
Valentijn, E. A., van der Werf, P. P., de Graauw, Th., de Jong,
T. 1996, A\&A 315, L145


\bibitem[\protect\citeauthoryear{V\"olk \& Xu}{1994}]{vx94} V\"olk, H.J., \& Xu, C. 1994, Infrared Physics and Technology, 35,
  527



\bibitem[\protect\citeauthoryear{Walsh et al.}{2002}]{wal02} Walsh, W., 
Beck, R., Thuma, G., et al. 2002, A\&A 388, 7

\bibitem[\protect\citeauthoryear{Wilke et al.}{2004}]{wil04} Wilke, K., Klaas, U., Lemke, D., Mattila, K., Stickel, M., et al. 2004, A\&A 414, 69

\bibitem[\protect\citeauthoryear{Wilke et al.}{2003}]{wil03} Wilke, K., Stickel, M., Haas, M., Herbstmeier, U., Klaas, U., et 
al. 2003, A\&A 401, 873

\bibitem[\protect\citeauthoryear{Wunderlich \& Klein}{1988}]{wk88} Wunderlich, E., Klein, U. 1988, A\&A 206, 47

\bibitem[\protect\citeauthoryear{Wunderlich et al.}{1987}]{wun87} Wunderlich, E., Wielebinski, R. \& Klein, U. 1987, A\&AS 69, 487


\bibitem[\protect\citeauthoryear{Xu \& Buat}{1995}]{xb95} Xu, C. \& Buat, V. 1995, A\&A 293, L65


\bibitem[\protect\citeauthoryear{Young \& Lo}{1997}]{yl97} Young, L.M. \& Lo, K.Y. 1997, ApJ 476, 131 
\end{thebibliography}
\end{document}